\documentclass[preprintnumbers, floatfix, preprintnumbers, letterpaper, superscriptaddress,nofootinbib, twocolumn]{revtex4}
\pdfoutput=1
\usepackage{graphicx}
\usepackage{microtype}
\usepackage{amsmath}
\usepackage{amssymb}
\usepackage{subfigure}
\usepackage{hyperref}
\usepackage{url}
\usepackage{xcolor}
\usepackage{color}
\usepackage{mathrsfs}
\usepackage{calrsfs}
\usepackage{amsfonts}
\usepackage{latexsym}
\usepackage{ragged2e}
\usepackage{epsfig}
\usepackage{textcomp}
\usepackage{phaistos}
\usepackage[utf8]{inputenc}
\makeatletter
\renewcommand\@makefnmark{\hbox{\@textsuperscript{\normalfont\color{purple}\@thefnmark}}}
\renewcommand\@makefntext[1]{%
  \parindent 1em\noindent
            \hb@xt@1.8em{%
                \hss\@textsuperscript{\normalfont\@thefnmark}}#1}
\makeatother

\usepackage{caption}
\DeclareCaptionJustification{justified}{\leftskip=0pt \rightskip=0pt \parfillskip=0pt plus 1fil}
\definecolor{vividviolet}{rgb}{0.62, 0.0, 1.0}
\definecolor{amaranth}{rgb}{0.9, 0.17, 0.31}
\definecolor{palatinateblue}{rgb}{0.15, 0.23, 0.89}
\definecolor{brightpink}{rgb}{1.0, 0.0, 0.5}
\definecolor{cornflowerblue}{rgb}{0.39, 0.58, 0.93}
\definecolor{deepcarminepink}{rgb}{0.94, 0.19, 0.22}
\definecolor{radicalred}{rgb}{1.0, 0.21, 0.37}

\hypersetup{ linktoc=all,
    colorlinks, linkcolor={palatinateblue},
    citecolor={brightpink}, urlcolor={amaranth}
}
\newcommand{\changeurlcolor}[1]{\hypersetup{urlcolor=#1}}
\graphicspath{{Images/}}

\renewcommand{\d}[1]{\ensuremath{\operatorname{d}\!{#1}}}

\graphicspath{{Images/}}
\renewcommand{\d}[1]{\ensuremath{\operatorname{d}\!{#1}}}
\makeatletter
\def\@fnsymbol#1{\ensuremath{\ifcase#1\or $\textleaf$ \or $\PHplaneTree$
\else\@ctrerr\fi}}%

\makeatother

\def\sideremark#1{\ifvmode\leavevmode\fi\vadjust{\vbox to0pt{\vss
 \hbox to 0pt{\hskip\hsize\hskip1em
 \vbox{\hsize1.5cm\tiny\raggedright\pretolerance10000
 \noindent #1\hfill}\hss}\vbox to8pt{\vfil}\vss}}}%
\begin{document}

\title{On Black Hole Thermodynamics, Singularity, and Gravitational Entropy}

\author{Yen Chin \surname{Ong}}
\email{ycong@yzu.edu.cn}
\affiliation{Center for Gravitation and Cosmology, College of Physical Science and Technology, Yangzhou University, \\180 Siwangting Road, Yangzhou City, Jiangsu Province  225002, China}
\affiliation{Shanghai Frontier Science Center for Gravitational Wave Detection, School of Aeronautics and Astronautics, Shanghai Jiao Tong University, Shanghai 200240, China}

\begin{abstract}
Black holes were found to possess properties that mirror ordinary thermodynamical systems in the landmark paper by Bardeen, Carter and Hawking almost half a century ago. Since then much progress has been made, but many fundamental issues remain. For example, what are the underlying degrees of freedom of a black hole horizon that give rise to said thermodynamical properties? Furthermore, classical black holes also harbor a spacetime singularity. Although it is often believed that quantum gravity would ``cure'' the singularity, as emphasized by Penrose, this viewpoint requires a deeper examination. In this review, I will examine the possibility that singularities remain in quantum gravity, the roles they may play, and the possible links between singularity and black hole thermodynamics. I will also discuss how -- inspired by Penrose's Weyl curvature hypothesis -- gravitational entropy for a black hole can be defined using curvature invariants, and the surprising implication that the entropy of black holes in different theories of gravity are different manifestations of spacetime curvature, i.e., their underlying microstructures could be different. Finally, I review the ``Hookean law'' recently established for singly rotating Myers-Perry black holes (including 4-dimensional Kerr black holes) that connect black hole fragmentation -- a consequence of the second law of black hole thermodynamics -- with the maximum ``Hookean force'', as well as with the thermodynamic geometry of Ruppeiner. This also suggests a new way to study black hole microstructures, and hints at the possibility that some black holes are beyond the Hookean regime (and thus have different microstructures). While examining the remarkable connections between black hole thermodynamics, spacetime singularities and cosmic censorship, as well as gravitational entropy, I shall point out some subtleties, provide some new thoughts, and raise some hard but fundamental questions, including whether black hole thermodynamics is really just ``ordinary thermodynamics'' or something quite different.

\begin{center}
{ \textit{An invited contribution to the SCRI21 Meeting ``Singularity Theorems, Causality, and All That: A Tribute to Roger Penrose''.}}
\end{center}
\end{abstract}

\maketitle

\section{Introduction: Black Hole Thermodynamics, Singularities, and Quantum Gravity}\label{1}

In Einstein's theory of general relativity (GR), a spacetime region can be sufficiently curved that nothing, not even light, can escape. Such regions, known as black holes, continue to fascinate both the general public and professional physicists. Black holes are certainly not an ``object'' in the physical sense. Instead the boundary of a black hole, its horizon, is only a non-tangible mathematical surface. Yet astrophysical observations have provided strong evidence that black holes do exist in our actual universe. From a theoretical physicist's point of view, there are at least two remarkable properties about black holes: (1) they behave like a thermodynamical system, namely they have entropy and temperature, despite being made of empty spacetime; (2) they harbor a singularity deep within. 

Now, property (1) is surprising because prior to the discovery of Hawking radiation \cite{hawking} no one expected a black hole to be able to radiate, and thus the correspondence with thermodynamics was initially taken as a mere analogy by Bardeen, Carter and Hawking. In fact, back then they referred to the laws of black hole \emph{mechanics} instead of \emph{thermodynamics} \cite{BCH}. Now we know that black holes have a well-defined Bekenstein-Hawking entropy \cite{b} and Hawking temperature. For an asymptotically flat Schwarzschild black hole of mass $M$ in 4-dimensions, the entropy $S$ and the temperature $T$ are given by
\begin{equation}\label{ST}
S=\frac{k_B c^3}{\hbar G}\frac{A}{4}, ~~\text{and} ~~ T=\frac{\hbar c^3}{8\pi G k_B M},
\end{equation}
where $k_B, c, G, \hbar$ denote the Boltzmann constant, the speed of light in vacuum, Newton's gravitational constant and the reduced Planck constant, respectively (hereinafter, they will be set to unity). Note that the entropy is proportional to the horizon area $A$, which is a very peculiar behavior compared to ordinary thermodynamical systems that we are more familiar with, such as that of a box of gas, whose entropy is proportional to the volume of the box. There had been investigations that tried to understand how such an area scaling of entropy could develop as a star approaches the black hole formation limit \cite{1810.01313,1910.13198}. However, this is not yet well-understood.

Crucially, Hawking proved that in classical GR, the area of the horizon cannot decrease, which indeed is what we expect once we have identified the horizon to be the measure of the entropy via Eq.(\ref{ST}). This is now referred to as Hawking's area theorem \cite{area}. Once quantum effects are included, black holes can shrink in size due to Hawking radiation, but the generalized second law \cite{GSL} (GSL) remains valid: that is, the entropy in the radiation together with the black hole entropy at a later time is always larger than the black hole entropy at the initial time before the radiation is emitted. Note the misnomer: there is really nothing ``generalized'' about the GSL, it is just the usual second law applied to the whole system; it is only generalized with respect to the original second law of black hole (only) thermodynamics. 

How the community came to gradually accept black holes as a \emph{bona fide} thermodynamical system has been explained in details by Wallace \cite{1710.02724,1710.02725} (see also \cite{1903.06276}). Here it suffices to emphasize that nowadays most theoretical physicists take black hole thermodynamics for granted, and a huge edifice of research is based on it. For example, the fact that Bekenstein-Hawking entropy is proportional to its surface area led to the ``holographic principle'' \cite{bousso} that the entropy contained inside an apparent horizon (even in cosmological settings) is bounded by the area of the horizon. This idea culminated in the ``gauge/gravity duality'' or ``AdS/CFT correspondence'' \cite{1501.00007} \footnote{In gauge/gravity duality, the physics of a system in the presence of gravity inside an asymptotically anti-de Sitter spacetime (``the bulk'') is equivalent to the physics of another non-gravitational field theory with one less spatial dimension at the conformal boundary. However, in most interesting applications, the field theory is not completely conformal. For example, when the field theory has a temperature dual to the Hawking radiation in the bulk, the temperature gives a length scale to the boundary. Hence the term ``gauge/gravity duality'' is more accurate than ``Anti-de Sitter/Conformal Field Theory (AdS/CFT) correspondence''. I have not much to say about this important aspect of theoretical physics, though there will be occasional related remarks.}
, which gave rise to an independent field of research altogether. 

Nevertheless, there are fundamental issues regarding black hole thermodynamics that require further examinations. One longstanding mystery is the underlying degrees of freedom or microstructures that give rise to the Bekenstein-Hawking entropy. What exactly does the entropy measure? Even after close to 50 years, we do not know why a static neutral Schwarzschild black hole -- the simplest of them all -- has the \emph{most} amount of entropy; why does spinning up a black hole \emph{decrease} its entropy? In fact, an avid reader has perhaps noticed that I chose to phrase property (1) as black holes behaving \emph{like} a thermodynamical system, instead of the more straightforward ``black holes are thermodynamical systems''. This is because there are some indications that black hole thermodynamics is \emph{not} ordinary thermodynamics despite the superficial resemblances. I will come back to this point in the final section.

Property (2) is the consequence of Penrose's singularity theorem \cite{1,1-2}, which earned him a Nobel Prize in 2020. Although mathematically speaking the theorem only proves geodesic incompleteness, physically black hole singularities are also where spacetime curvature becomes unbounded. The divergence renders GR useless in describing the physics at singularities. One should of course be very suspicious when calculations yield infinity. Thus it is suspected that the diverging curvature is a signal that GR is no longer applicable near the singularity. From the point of view of treating GR as an effective field theory \cite{9512024,0311082}, this makes sense -- a sufficiently high curvature corresponds to high energy scale beyond some ultraviolet cutoff that GR can no longer be trusted (even making this statement precise is nontrivial because the relationship between ``high curvature'' and ``high energy'' is not so straightforward. Among other reasons, one crucial difficulty is the notion of ``gravitational energy'', itself not well understood; see the recent article \cite{2204.03869}). Therefore, it is usually suspected that once quantum theory is taken into account properly, the singularity would be ``cured'', i.e., rendered non-singular. In other words, the classical singularity is often viewed as something that is unwanted and which is not to be taken as ``real'', merely a signal that GR is no longer the valid theory. However, in view of the fact that we do not really have a fully working theory of quantum gravity (QG), we should keep our options open and consider seriously the possibility that some kind of singularities might persist down to the QG regime. It might be replaced by something else, a ``quantum singularity'' that is perhaps less pathological than its classical counterpart, yet nevertheless might not be completely cured. (It could also be possible that whether a singularity is resolved in QG depends on its ``strength'' \cite{1606.04932}.) In the subsequent discussions, we shall see why it might be important to keep some singularities around. In fact, they could be a feature, rather than a bug, of general relativity.

We also recall that the mainstream view of the end state of Hawking evaporation is that eventually the black hole completely disappears, and one is left with an empty spacetime\footnote{An alternative view that some sort of remnant persists is not entirely ruled out \cite{1412.8366}, though they are widely argued to be problematic. Indeed, the recently popular weak gravity conjecture \cite{0601001,2201.08380} was partly formulated to prevent such a state from ever arising.}, though if one considers the quantum vacuum before black hole formation and after the evaporation they might be different, as proposed by the soft hair picture \cite{1601.00921,1706.07143}. This, at least to me, seems to suggest that it is implicitly assumed that either the singularity was never really there (not even a quantum version), or that the quantum singularity evolves such that at the end it dissolves completely, and so we do not have a naked quantum singularity\footnote{Here we recall Penrose's statement to the contrary \cite{penrose1999}: ``It is hard to avoid the conclusion that
the endpoint of the Hawking evaporation of a black hole would be a
naked singularity – or at least something that one [sic] a classical
scale would closely resemble a naked singularity.'' In fact, Russo has argued that a complete evaporation would lead to a catastrophic event at the end of the evaporation \cite{russo} -- an outbrust of Planckian curvature wave similar to a ``thunderbolt'' singularity \cite{penrose1999, thunder1, thunder2}.}. If the second possibility holds, then it is plausible that quantum singularity is somehow tied to the horizon (which also needs to be replaced by a quantum notion thereof) so that the singularity is guaranteed to dissolve when the horizon does, i.e. some version of quantum cosmic censorship continues to hold\footnote{In fact, cosmic censorship is not a strictly classical phenomenon, for under some situations one needs quantum effects to maintain the censorship \cite{0011003,0810.0079} that would have otherwise failed. See also \cite{1608.05366}.}. 
The mainstream viewpoint regarding Hawking evaporation is therefore quite consistent with the idea that cosmic censorship remains relevant in QG. In the next section I will continue to discuss the possibility that the singularity is somehow related to the horizon. Granted that this is not a mainstream view, and I myself am not entirely convinced of it anyway, I think it is important to explore such a possibility. 

The structure of this review -- which is not meant to be comprehensive -- is as follows: in the next section I will discuss the roles singularities might play, and why they might continue to exist in some ways in QG. 
In Sec. (\ref{3}) I will discuss the idea of gravitational entropy, and a proposal of gravitational entropy based on my recent work with Gregoris \cite{2109.11968}, which was motivated by Penrose's Weyl curvature hypothesis (in the cosmological context of the ``arrow of time''). This approach suggests that different constructions of curvature invariants might account for black hole entropy in different theories of gravity, which in turn might imply that non-GR black holes have different microstructures.   
While the main objective is to shed some light on the nature of Bekenstein-Hawking entropy, the bigger theme that underlies this investigation concerns the hope that one might gain a better understanding of gravitational entropy in more general contexts (such as gravitational waves), and the role singularities and cosmic censorship might play in this. 
In Sec. (\ref{4}), I will review the recently established ``Hookean law'', based on a collaboration with Di Gennaro and Good \cite{2108.13435}, that can be viewed as a new aspect of black hole thermodynamics. In short, a Hookean law ``$F=kx$'' analogous to a spring is defined for black holes, whose maximum value corresponds to black hole fragmentation (a black hole splitting into two as the latter configuration has a higher entropy). Surprisingly this can be related to the thermodynamic geometry of Ruppeiner, as well as to the ``maximum force conjecture'' in GR \cite{0210109, 1408.1820, 0607090, 724159}. We also suggested in \cite{2108.13435} that there might be non-GR black holes that go beyond the Hookean regime, and thus possess different microstructures, as also conjectured in Sec. (\ref{3}). Finally I conclude with some prospects and suggestions for future investigations in Sec. (\ref{4}).

\section{The Roles of Singularities and Why They Might Persist in Quantum Gravity}\label{2}

Not all singularities are the same -- some are worse than others. The problematic ones are timelike singularities, which are singularities \emph{in space}, such as the ones inside the rotating Kerr black hole, or the charged Reissner-Nordsrt\"om black hole. They would be visible to anyone that enters the black hole, and worse still, if the horizon is somehow destroyed, be visible to the rest of the universe. 
This is problematic because we lack a theory that can properly describe the physics of singularity, and if a singularity can be observed, it can also affect physics far away. General relativity would then lose its predictive power, since given an initial condition we can no longer evolve a system in time without also knowing what sort of boundary condition we should prescribe to the singularity (i.e., Cauchy problem is no longer well-posed). This is the reason behind Penrose's proposal of cosmic censorship conjecture \cite{1-2}. The conjecture comes in two forms: the weak form essentially forbids naked singularity so that the rest of the universe outside black holes is predictive, whereas the strong form requires this to be true also for the interior observers, namely we should not expect a stable timelike singularity to be physical. 
In this work, by cosmic censorship I mean the weak version. The strong version will be explicitly referred to as such.

A spacelike singularity, which is ``located in time'', like the one inside a Schwarzschild black hole, is harmless because being ``the end of time'' it cannot affect anything. Incidentally, the big bang singularity which is ``the beginning of time'' is also harmless because even though we still cannot fathom the physics right after the Big Bang, the subsequent evolution of the universe is relatively well-understood and we can just take an initial slice at some later time and evolve that forward. Lastly, a null singularity, which can occur on the horizon of some black holes, can be problematic if it has a diverging curvature, since then its neighborhood can have unbounded curvature that can affect the exterior universe. I have recently reviewed various aspects of weak cosmic censorship \cite{2005.07032}, which I shall refer the readers to for more details. 

In this section I wish to focus on one issue: why is cosmic censorship relevant at all? Especially if we think that there is no such thing as an actual singularity (if it is cured by QG -- not surprisingly there are different attitudes among physicists towards singularities in relation to QG \cite{2112.08531}), then why should we even be bothered by cosmic censorship? Well, because Nature seems to care. The fact is that we have not seen any naked singularity (or whatever replacing it in QG). On the contrary, astronomical observations reveal numerous black holes that are rotating very close to the extremal limit \cite{1903.11704}, which is as fast as cosmic censorship would allow. In addition, theoretical attempts to destroy black hole horizons reveal how difficult such tasks can be. In other words, the laws of physics seem to abhor naked singularities. That is to say, it does seem that cosmic censorship is generically true (quantifying the meaning of ``generic'' is of course not easy). It seems strange that Nature would try so hard to uphold cosmic censorship if there is nothing worrying inside black holes\footnote{We can further compare this with the ultraviolet catastrophe (Rayleigh–Jeans catastrophe) in classical thermodynamics. In that case, a divergence was predicted by the classical theory, which of course did not match observations. Once quantum mechanics is taken into account, the infinity is ``cured'' and one obtains the Planck spectrum. Thus quantum mechanics is required to explain the observed spectrum. In our context, however, the classical theory of GR \emph{already} predicts the correct observation: namely there is a censorship bound, without which one runs into the problem with naked singularity. If QG cures away the singularity, it also removes the need for cosmic censorship. So quantum modifications in this case, seemingly introduced a new question instead of solving an existing one.}. This is of course not even close to a serious argument. Indeed, this suggestive opinion has at least one loop hole: there are \emph{regular black holes} beyond GR that do not have a singularity by construction, yet they typically have an extremal configuration. Though we do not call it by the name cosmic censorship, \emph{it is rather the same phenomenon}: there are values of the parameters that the configuration ceases to have a horizon. I will come back to discuss regular black holes later.

For now, let us take a step back and recall the usual assumptions of the singularity theorem, and ask if we still expect singularities at the quantum level. The original singularity of Penrose, and many versions thereafter, assume some kind of energy conditions \cite{jose}. 
Even the weakest classical energy condition, the null energy condition (NEC), is known to be violated once quantum effects are considered.
It is possible to prove singularity theorems assuming only the averaged energy conditions \cite{1012.6038,1907.13604,2012.11569,2108.12668}, integrated over some regions instead of being valid pointwise, which are believed to hold at the semi-classical level. 
(Even taking into account Hawking evaporation, which of course violates NEC, a singularity theorem can still be proved \cite{1909.07348}.)
But what happens in the QG regime? One naturally suspects that even the averaged energy conditions may cease to hold. One is faced with the hard problem of deciding which principles should we still hold on to entering the realm of QG. Should we keep some kind of energy condition? Should we even keep Lorentz invariance\footnote{Interestingly, and somewhat surprisingly, there may be nontrivial connections between Lorentz symmetry and the second law of thermodynamics \cite{0603158,0702124,0811.0943}, and local Lorentz symmetry may be an emergent property of the macroscopic world with origins in a microscopic second law of causal horizon thermodynamics \cite{0804.2720}.}, for example? Or some properties from quantum information theory, perhaps? With the lack of experimental guidance, it is often down to personal bias as to what should be regarded as fundamental. 

One possibility is to hold on to some notion of thermodynamics (or inspired from it), in particular the second law. As Eddington has famously declared \cite{eddington}, ``\emph{The law that entropy always increases holds, I think, the supreme position among the laws of Nature. [...]  if your theory is found to be against the Second Law of Thermodynamics I can give you no hope.}''
The second law reflects some sort of order -- it is the statement that universe has an arrow of time \cite{Price3}. Is the second law, or at least the notion of entropy, ``sacred'' enough to survive in QG? One might argue that microscopic physics experiences no arrow of time because the laws of physics is fundamentally time symmetric. Indeed, the entropy of the configuration of a gas in a box tends to increase over time (molecules released from a bottle of perfume at a corner of the box quickly spread uniformly throughout). If we take a video recording of the process and play it backward it would be obvious that the footage is played backward. However, if we zoom in to record the individual motion of gas molecules and their collisions, the footage would appear essentially the same whether we play it forward or backward in time. For this reason we might be tempted to suspect that at the QG level the arrow of time -- which is in this sense emergent and not fundamental -- is not important. However, the Big Bang was in the QG regime, and the universe \emph{did} evolve forward in time, giving rise to the macroscopic arrow of time. Therefore it is not so clear that QG does not distinguish between some kind of low entropic state from some other high entropic state (this is also related to the difference between an initial singularity and a final singularity, which I will discuss below). Let us therefore, at least entertain the possibility that the notion of entropy, and some kind of second law, remains valid in QG. Indeed, assuming that the GSL holds and with some machinery from quantum information theory, Wall \cite{1010.5513} was able to prove a version of the singularity theorem. Thus, as per our discussion above, it could be possible that a similar result may hold in QG.

While there is no concrete evidence that supports the notion of singularity in full QG, there is at least one evidence that cosmic censorship remains relevant at least in the low energy limit of string theory (not just semi-classical GR). The case in point is the evaporation of the charged dilaton black hole (the ``GHS'' solution \cite{ghs, g, gm}). As Hiscock and Weems have shown \cite{HW}, the Hawking evaporation of charged black holes is highly nontrivial, but the bottom line is this: for a Reissner-Nordstr\"om black hole, if the initial charge-to-mass ratio is sufficiently large, then the black hole simply steadily discharges towards the Schwarzschild limit. However, if it started with a low charge-to-mass ratio, it will radiate predominantly neutral particles and so the charge-to-mass ratio first increases, though eventually it will start to decrease. The extremal configuration is therefore not only unachievable (which is to be expected from the third law of black hole thermodynamics since it has zero Hawking temperature), but that any black hole that comes close to the extremal state will be ``repelled'' away. Note that this is also the expectation from the weak gravity conjecture. 
Now, we can naively apply the Hiscock-Weems method to study the evolution of the evaporating GHS black hole, which can be understood as a string theoretic version of charged black holes, at least in the low energy limit. In this case we found that the black hole continues to approach the extremal limit without turning around. This violates the spirit of cosmic censorship because the extremal GHS black hole is a null singularity, with a nonzero Hawking temperature and a diverging curvature. Therefore, even if the extremal state is not exactly attained, the theory becomes non-predictive when the black hole gets arbitrarily close to being extremal (and so has arbitrarily large curvature). The point is that it does not take an exact singular state with infinite curvature to be problematic, an arbitrarily large one is equally bad.  We can turn the situation around and ask, if we \emph{impose} cosmic censorship, i.e., if we require that the evolution of the charge-to-mass ratio to eventually turn around, how do we go about implementing it? The Hiscock-Weems model involves the charge loss rate due to particle production via the Schwinger effect\footnote{Note that the Schwinger effect is necessarily suppressed for all values of black hole parameters \cite{gibbons}, but it is crucial that the term be included, otherwise all Reissner-Nordstr\"om black holes will tend to extremality, and potentially violates cosmic censorship \cite{2102.05519}.}, which should receive correction in the presence of the dilaton. In fact, by assuming a rather simple minimal ansatz for the correction, such that the charge-to-mass ratio turns around, Yao and I found that we could re-derive the correct charge particle production rate in the dilaton-Maxwell theory \cite{1907.07490}, which was previously obtained in the literature via a direct QFT computation \cite{1305.2564}. 
Explicitly, the ansatz we assumed for the exponential term in the Schwinger process takes the form:
\begin{flalign}
\exp\left(-\frac{r_+^2}{Q_0Q} + \mathcal{C}\frac{Q}{Q_0}\right) = &\exp\left(\frac{\sqrt{2}M (\mathcal{C}-2)}{Q_0}\right) \\ \notag &\times \left[1-\frac{2+\mathcal{C}}{Q_0}\varepsilon + O(\varepsilon^2)\right],
\end{flalign}
where $Q_0:=e/(\pi m^2)$, and $\varepsilon$ is the deviation away from the extremal charge configuration $Q=\sqrt{2}M-\varepsilon$, while $\mathcal{C}$ is the correction term due to the presence of the dilaton. That is to say, $\mathcal{C}=0$ corresponds to the usual exponential term in the Schwinger effect in standard quantum electrodynamics. The additional dimensionless term $\mathcal{C}{Q}/{Q_0}$ essentially renders the Schwinger effect \emph{less} suppressed. How much of enhancement in the production rate do we need to discharge the black hole away from extremality? It turned out that for any fixed value
of $0 < \mathcal{C} < 2$, we can always choose $M$ sufficiently large so that there are initial datum that would lead to extremality
under Hawking evaporation and Schwinger pair production. Therefore, if we want to impose cosmic censorship as per the discussion above, the minimal choice is $\mathcal{C} = 2$, which is indeed the correct value obtained via QFT.
This example shows that cosmic censorship can be used as a guiding principle to deduce other physics. Furthermore, this example provides an indirect, but strong, evidence that cosmic censorship remains relevant at least in the low energy limit of quantum gravity. (Di Gennaro and I also conjectured in \cite{2103.05516} that as any black hole undergoes
Hawking evaporation, its parameters should evolve in such a way that it avoids becoming extremal, not just avoid becoming a truly naked singularity. This could be used as a guide to rule out non-standard models of Hawking evaporation.) In addition to the Maxwell-dilaton case, recently some authors have investigated black hole evaporation in the presence of higher order curvature terms (with or without a dilaton), such as the Gauss-Bonnet term. It was found that cosmic censorship can be violated due to the formation of either a naked singularity or a high-curvature elliptic region \cite{2103.00257,2205.13006,2205.13007,2207.10692}.

Having discussed the various hints that singularities might remain in QG, let us now turn to discuss what roles they may play. That is, there might be good reasons why Nature wants to keep singularities around. 
First of all, recall that the Schwarzschild metric admits a negative mass solution. While mathematically valid, it is unphysical, otherwise we would not have a stable ground state in GR. The positive energy theorem guarantees stability in this sense. The negative mass Schwarzschild solution is however, not a black hole, because it has no horizon. Instead it is a naked singularity, which would be obvious if we draw its Penrose diagram. The interior solution of a Reissner-Nordstr\"om or a Kerr black hole inside the inner horizon (ignoring the stability issue for now) is actually similar, which also can be seen from their Penrose diagrams. With respect to an interior observer, the timelike singularities within these black holes are naked (the strong cosmic censorship conjecture wants to rule out such a scenario). Indeed, we see the ``negative mass'' phenomenon here as well: gravity becomes repulsive near these singularities \cite{2005.07032}, and freely falling observers along geodesics (of a neutral particle in the charged case), cannot hit the singularities. Instead they are repulsed away into another universe, if we take the maximally extended solutions at face value (which we should not, because of the instability of the inner horizon).
Therefore, one can say that even though a naked singularity has a negative mass and is thus not allowed, it is nevertheless acceptable to have such a singularity hidden behind the horizon (if we only accept the weak cosmic censorship instead of the strong version\footnote{It is worth emphasizing again at this point that singularities are problematic only when they cause the theory to lose predictability. If the instability of the inner horizon turns into a spacelike singularity, there is actually no problem with predictability, and as such, is a good thing. Many physicists, however, are uncomfortable with singularities, even if they are spacelike.}) -- the mass as measured from outside is positive. It has therefore been suggested that singularities play the role of ensuring stability even in QG. Namely, if QG corrections render a classical singularity regular, there would no longer be any need for cosmic censorship, and in principle, the ``regularized singularity'' can therefore be naked. But then there is no clear reason why the energy of these regularized singularity should be bounded from below; they can be very negative. In other words, cosmic censorship could therefore play the role of ruling out unphysical solutions that would otherwise render the theory unstable \cite{9503062}.  

From the above discussion, we already observed a curious relationship between the singularity and the horizon, they seem to conspire to give a positive mass. In fact, there seems to be more to this story. In an asymptotically flat spacetime we could compute what is the mass of the black hole in the usual way. This is the Arnowitt-Deser-Misner (ADM) mass --  which is essentially the mass we would physically assign to the black hole. On the other hand, there is another mass, or rather ``internal energy'' $E$, that is physical in the sense that it is the energy that appears in the first law of black hole thermodynamics: $\d E=T\d S$. There is no obvious reason why we should have $E=M$. However, this is a robust feature in GR. For example, in anti-de Sitter spacetime, the rotating Kerr-AdS black hole has a mass that is analogous to the ADM mass (the Abbott-Deser mass) \cite{AD, 0506057}, which is \emph{exactly} the same mass that is required for the first law to hold \cite{1506.01248}. However, this nice property that the gravitational mass is the same as the thermodynamic mass need not hold for regular black holes \cite{1411.0833, 2103.14413}, in which by construction there is no singularity within. While the gravitational mass is  positive, it is not the same as the thermodynamic mass in the first law (alternatively we could give up the Bekenstein-Hawking area law for these black holes \cite{1411.0833}). Thus, we see that the singularity is inherently related to black hole thermodynamics -- which are usually thought to be the properties due solely to the horizon. 

Let us make a few remarks on regular black holes \cite{2208.12713}. These are ad hoc geometries in which the singularity is somehow regularized by some exotic matter field or by some quantum effects. Regular black holes can have multiple horizons, which means that they can become extremal much like a Reissner-Nordstr\"om black hole. A Reissner-Nordstr\"om black hole can lose its horizon and expose the singularity if the charge becomes too large. The cosmic censorship conjecture says that this should not happen. The censorship bound is therefore a mathematical condition that requires the existence of a horizon. Such kind of bound also exists for regular black holes with multiple horizons, despite they harbor no singularity! In general, therefore, one should not equate the censorship bound with the horizon existence condition, as there is nothing to be censored in a regular black hole. However, the fact that these black holes have an inner horizon, the same sort of arguments leading to mass inflation inside a Reissner-Nordstr\"om black hole also suggests that we should not trust the validity of their interior solution. But it is precisely the interior solution that is of interest in this context. If, under mass inflation, the inner horizon turns into a singularity, then this would defeat the entire purpose of constructing a regular black hole solution in the first place (see, however, \cite{2209.10612}). This conundrum has been discussed in details by Carballo-Rubio et al. \cite{1805.02675, 2101.05006}, the bottom line is that it is far from clear whether multi-horizon regular black holes are self-consistent. (If there is only one horizon, there is also no ``horizon existence condition'' -- the solution always exists.) There are some solutions that are claimed to be stable \cite{2205.13556, 2207.08864}, but to prove this at the mathematically rigorous level would require more work. Their thermodynamics should also be further examined. 

This is a good place to mention some rigorous results regarding the inner horizon. The naive Penrose diagram of Reissner-Nordstr\"om and Kerr black holes found in the standard GR textbooks are now known to be incorrect. The Cauchy horizon is itself essentially singular (``weak null singularity'' \cite{1311.4970,1710.01722}). Its instability does eventually lead to the formation of a singularity, co-existing with a portion of a Cauchy horizon \cite{1704.05790,1710.01722,1912.10890,2001.11156,2105.04604}. This singularity is most likely spacelike. Whether this happens for the regular black holes requires a proof at the same level of rigor. Of course, at some point, quantum corrections will change the classical picture. However, even at the quantum level, it is likely that Cauchy horizon would be inherently singular in some ways \cite{1005.2999}. In a recent work \cite{2206.07001}, Bousso and Shahbazi-Moghaddam even argued that ``quantum singularities blurs the line between an ordinary singularities and a Cauchy horizon.'' Remarkably, whether spacetime continues to exist beyond a Cauchy horizon may be probed by the thermal one-point function of a massive field \emph{outside} the black hole \cite{2011.01004, 2206.00198}.

\section{Gravitational Entropy, the Second Law, Gravitational Waves and Singularities}\label{3}

I have briefly mentioned the arrow of time problem in the previous section. As Penrose pointed out \cite{penrosetime}, the Big Bang is special because entropy was very small back then, compared to what it could have been. More specifically Penrose computed the full phase space volume, and the volume that corresponds to the actual initial conditions of our observable universe. The ratio is about $10^{10^{123}}$ \cite{123}, dubbed the ``Penrose number''\footnote{The estimate made by Penrose is essentially based on the number of microstates corresponding to the maximum possible entropy, which is given by the Bekenstein-Hawking entropy of a black hole with the total mass of the observable universe. Surprisingly, this number is also close to the one obtained by assuming we are living inside a de Sitter spacetime (now that we know that the universe undergoes an accelerated expansion), in which the de Sitter horizon entropy is $S\sim 1/\Lambda \sim 10^{122}$ in the Planck units. The current \emph{actual} entropy inside the observable universe is about $10^{104}$, which also includes contributions from supermassive black holes \cite{0909.3983}. 
This is an example of the ``holographic principle'' in action, that the entropy of the cosmic horizon bounds the interior entropy.} in \cite{0711.1656} by McInnes, who also pointed out that the number of internally consistent universes in the string landscape -- usually cited as $10^{500}$ for definiteness -- is utterly negligible by comparison. In other words, it appears that it is extremely improbable that our universe started in the way it did. Explaining this is the arrow of time problem, which involves some very tricky issues (including the infamous ``measure problem'' \cite{0609095} of how to make sense of probability when you have an eternal inflation), and deserves another review by itself, therefore it is not our aim to discuss it in depths here. For our purpose, we shall focus on only one aspect: namely gravitational field has entropy, which we still do not know how to properly quantify. This is important for the arrow of time problem, as one should track all sorts of entropy -- not just the matter degrees of freedom -- to say whether entropy is high or low. 

From the uniform and Planck spectrum of the cosmic microwave background radiation, we see that the matter sector did attain thermal equilibrium in early times. Yet we know that the overall entropy of the universe keeps on increasing (thus giving rise to the arrow of time). Therefore, Penrose realized that the initial entropy must be low in the sense that gravitational entropy was low. 
Although we do not know much about gravitational entropy, it makes sense to suspect that it should be somehow related to the Weyl curvature. This is because in GR, it is not the full Riemann curvature tensor that appears in the Einstein field equations. Only the trace part of the Riemann tensor -- the Ricci tensor (and its contraction, the Ricci scalar) is related to the matter-energy stress tensor. 
The traceless part of the Riemann tensor is exactly the Weyl tensor, whose components are denoted by $C_{abcd}$. In other words, the Weyl curvature is the only part of the curvature that exists in free space
in the absence of matter, and thus has to do with gravitational degrees of freedom only. Penrose therefore proposed the Weyl curvature hypothesis \cite{wcc1}, which states that the initial Big Bang entropy must correspond to a very small Weyl curvature (which measures ``tidal deformation'' in GR), whereas the Weyl curvature is expected to be large at a putative Big Crunch, or at the black hole singularities. The different characters of the initial and final singularities may indicate that their resolutions (if any) in QG may need to be different. Quite independently of the arrow of time problem, it is an interesting question by itself to properly understand gravitational entropy. Mostly we are only confident of one special case: the Bekenstein-Hawking entropy of a black hole. Before coming back to the question of what black hole entropy really measures, let us comment on the bigger picture.

As mentioned in the Introduction, Hawking's area theorem \cite{area} proved that classically\footnote{A strengthened version of the theorem can be proved even when the NEC is somewhat violated \cite{1711.06480}. At present it is not clear whether the relaxed energy conditions that guarantee the validity of a semi-classical version of the area theorem would also be the same conditions that forbid Hawking radiation and vice versa.} the area (and hence the entropy) of a black hole horizon cannot decrease. Recently, the area law has even been ``tested'' using the first ever observed gravitational wave data (of course, as a mathematical result, the area \emph{theorem} needs no ``test'' -- the real purpose of such test is to see if the \emph{assumptions} of the theorem hold in the actual universe) \cite{2012.04486}.

In view of the current huge interests in gravitational wave physics, it is intriguing to note that Hawking's original motivation was to study how much energy can be emitted by gravitational waves during black hole mergers. The area theorem implies that the area of the post-merger black hole must be at least as large as the sum of the areas of the initially separate black holes. This in turn gives an upper bound on how much of black hole mass can be converted into gravitational waves\footnote{A lower bound on
the entropy of a black hole can be deduced from a bound on the minimal (Tolman) redshift
factor of gravitational waves emerging from the vicinity of its horizon \cite{1811.12283}.}. It would be very interesting to further understand this process from the point of thermodynamics. Recall that a black hole that undergoes Hawking evaporation still satisfies the GSL since the entropy in the Hawking particles compensates the reduction in the Bekenstein-Hawking entropy. So why is it that gravitational waves cannot compensate for a putative decrease in the post-merger area? The area theorem amounts to an indirect argument that such a process is not possible. It must be that gravitational waves cannot carry so much entropy. To understand why not we must be able to quantify gravitational entropy in general, not just that of black holes, but also (at least) for gravitational waves \cite{smolin}. Such a proposal has been made by Clifton, Ellis, and Tavakol, utilizing the Bel-Robinson tensor, which is constructed from the Weyl tensor and its dual, and has a natural interpretation as the effective ``super-energy-momentum tensor'' of free gravitational fields \cite{1303.5612}, yet much is still to be understood\footnote{The Clifton-Ellis-Tavakol gravitational entropy was also shown to be increasing during structure formation in Szekeres Class I models \cite{2205.02985}. Interestingly, in the presence of shear, the  Clifton-Ellis-Tavakol gravitational entropy can increase even though the Weyl curvature \emph{decreases}. Thus, the Weyl curvature hypothesis is not strictly valid in all cosmological spacetimes \cite{2004.10222}.}. 

We may even consider the extreme case to try to see why such a process is impossible: consider two black holes collide and completely ``annihilate'' into gravitational waves. It turns out there is a good reason to prohibit this -- the end result would not just be an empty space plus gravitational waves. Rather, it would likely create a naked singularity and thus violates cosmic censorship. For the same reason an asymptotically flat and 4-dimensional black hole cannot split into two. This is Theorem 12.2.1 in Wald's GR textbook \cite{Wald}. We note that, as pointed out in \cite{2201.01939}, the premise of the theorem is to assume that spacetime is ``predictable'', i.e., no unknown information coming in from a putative naked singularity. Thus the contrapositive statement would be that if black holes do bifurcate, the spacetime would contain a naked singularity (or perhaps other anomalies). The theorem however \emph{assumes} an arrow of time \cite{2201.01939} -- otherwise the same theorem would ``prove'' that black holes cannot merge or cannot form from the collisions of gravitational waves. In other words, \emph{if} we know which direction is the future (i.e. the direction entropy increases), and \emph{if} cosmic censorship holds, then it follows that black holes cannot bifurcate or otherwise reduce their total areas\footnote{It is debatable whether this is really a problem. For example, when singularity does form when a black string or black ring undergoes ``pinch-off'' under Gregory-Laflamme instability \cite{1107.5821}, the loss of classical predictability of the system is argued to be quite small. This is also the case for the naked singularity formation under
black hole collision in higher dimensions. To quote \cite{1812.05017}, ``even if cosmic censorship is violated, its spirit remains unchallenged.'' Is a little violation acceptable?}. So again the arrow of time creeps into the problem; to understand the thermodynamical aspect of gravitational wave we need to have a good understanding of gravitational entropy, not just for black holes but for gravitational waves, and possibly also for singularities (see the final section for more comments on this).

Even in the case of black hole entropy -- the best understood of all instances of gravitational entropy, it is still unclear what is the underlying microstructures or microstates that this entropy supposdly measures. This question can be asked on different levels. It would certainly be best to understand this at the full QG level. One approach from string theory (Strominger-Vafa \cite{9601029}) was able to derive the Bekenstein-Hawking entropy by counting the microstates made of a stack of branes that collapse to form an extremal black hole when some coupling parameter is tuned. It is, however, unclear if this can be done for generic black holes. Incidentally, one interesting aspect is that the end result of the ``collapse'' still has the same entropy as the initial configurations. In other words, the initial states that form the black hole saturates the Bekenstein-Hawking entropy. This is non-generic: a typical collapse of a stars to form a black hole, for example, involves a huge increase of entropy by a factor of $10^{20}(M/M_{\odot})^{1/2}$, where $M_{\odot}$ denotes a solar mass \cite{hara,2206.11870}. While specific examples can be constructed at the classical level whose initial conditions do saturate the Bekenstein-Hawking entropy bound, such configurations are highly nontrivial and finely tuned \cite{1611.04044}. This calls for a better understanding of two challenges: (1) Why does gravitational collapse generically involve such a huge increase in entropy? (2) Notwithstanding the Strominger-Vafa approach (and others) at the QG level, can we understand, from the point of view of Weyl curvature, how to interpret the Bekenstein-Hawking entropy? 

One naive but straightforward proposal to address (2) was proposed in \cite{epjcbh1}. The idea is simply to treat the contraction $C_{abcd} C^{abcd}$ as the entropy density and integrate over the appropriate spacetime region to obtain the entropy of the gravitational field as
\begin{equation}
\label{prop}
S_{\rm grav}:=\int C_{abcd} C^{abcd} \d V_4\,,
\end{equation}
where $\d V_4=\sqrt{h}~ \text{d}^4x$ is the hypersurface volume element, which after integrating from $r=0$ to the horizon, reproduces the Bekenstein-Hawking entropy up to a constant prefactor.  
Due to dimensionality reason, this only makes sense in 5-dimensions, unless we modify the proposal in an ad hoc way by raising the integrand to the appropriate dimension-dependent power.
Another problem with this proposal is the need to set a cutoff in the lower limit of the integral to avoid a divergence at the singularity $r=0$.

In \cite{2109.11968}, Gregoris and I proposed a different construction. 
First, we note that for Petrov type D spacetimes, $\Psi_2=C_{abcd} n^a m^b \bar{m}^c l^d$ is the only nonzero Weyl scalar. For a static spherically symmetric black hole in 4-dimensions (no assumption of asymptotic flatness) with a metric component $-g_{tt}=f(r)$, we have, upon adopting the null coframe,
\begin{eqnarray}
\label{weyl1}
\Psi_2=\frac{r^2 f'' -2rf' +2f -2}{12 r^2}\,,\\D W=\frac{\sqrt{2 f}(r^2 f'' -2rf' +2f -2)}{8 r^3}\,,
\end{eqnarray}
where $DW$ is the component of the first order frame derivative of the Weyl tensor $W=C_{abcd}$; $D:=n^a\nabla_a$ being the Newman-Penrose directional derivative. The Bekenstein-Hawking entropy can then be reproduced by
\begin{equation}
\label{entropyformula}
S_{\rm grav}:=\frac{1}{3 \sqrt{2}}\int_0^{r_H} \int_{S^2} \Big| \frac{D W}{\Psi_2}\Big| \frac{r^2 \sin \theta }{\sqrt{f(r)}} \d r \d\theta \d\phi =\pi r_H^2=\frac{A}{4}\,,
\end{equation}
without a need of any cutoff near the singularity.  Adopting the same language of  Clifton-Ellis-Tavakol \cite{1303.5612} by writing
\begin{equation}
S_{\rm grav}= \int_V \frac{\rho_{\rm grav}}{T_{\rm grav}}\d V \,,
\end{equation}
we can interpret the Cartan invariants $DW$ and $W=|\Psi_2|$  to be the ``energy density and temperature of the gravitational field", respectively. 
The quantity $DW$ is also known as the Cartan invariant, which vanishes on the horizon (and thus serves as a local ``detector'' of a horizon \cite{cartan}). Unlike the previous proposal in Eq.(\ref{prop}), $S_{\rm grav}$ is finite because of the function $1/\sqrt{f(r)}$ entering the hyperspace volume element $\d V_3=({r^2 \sin \theta}/{\sqrt{f(r)}}) \d r \d\theta \d\phi$, which cancels the divergence behavior of the integrand $\rho_{\rm grav}/{T_{\rm grav}}$ at the horizon. The prefactor ${1}/{3 \sqrt{2}}$ in proposal (\ref{entropyformula}) is chosen \emph{a posteriori} to match the Bekenstein-Hawking entropy, which is still a shortcoming (that is also present in \cite{1303.5612} and \cite{epjcbh1}). 

Our proposal also works in the cosmological setting. Generalizations to higher dimensions, though not trivial, should be possible. We studied the 5-dimensional case explicitly as well \cite{2109.11968}.  Rather surprisingly, it also works for Reissner-Nordstr\"om spacetime (whereas this is not the case for proposal (\ref{prop})). This seems to suggest that the Bekenstein-Hawking entropy of a Reissner-Nordstr\"om black hole \emph{only} counts the gravitational degrees of freedom. 

I think there is one major mystery yet to be understood in these proposals: the integral from $r=0$ up to the horizon is an integral \emph{in time} for the Schwarzschild case (since the vector field $\partial_r$ is timelike in the interior), whereas for the cosmological spacetimes an integral out towards the cosmological horizon is an integral in space (granted that space itself is expanding hence there is a time dependence), and worse still for the Reissner-Nordstr\"om case, part of the integrals involve spatial integrals whereas the region between the two horizons is a temporal integral. Yet these distinctions do not seem to matter in calculating $S_\text{grav}$, \emph{why is that}? Naively one might suspect that entropy has to do with the temporal direction (in which the entropy increases -- one such proposal for black hole entropy was made in \cite{1010.5844}), but this does not appear to be the case.

Another interesting aspect is that our proposal is purely geometrical, so it will \emph{always} give an area law regardless of the underlying gravitational theory. However, we know that not all black holes in modified gravity satisfy the simple area law. For example, $S \neq A/4$ in general in $f(R)$ gravity \cite{entropyf1,entropyf2,entropyf3,entropyf4,entropyf5,emre}, the Lovelock theory \cite{0311240}, and the Einstein-Gauss-Bonnet gravity \cite{gbentropy1,gbentropy2,gbentropy3}. 
This suggests that some non-GR black holes might have different underlying microstates or microstructures from black holes in GR, in the sense that their gravitational entropy is not given by the integrand proportional to $DW/\Psi_2$. In the next section, I will review another line of thought that also suggests this conclusion.

\section{The Hookean Law: A New Aspect of Black Hole Thermodynamics}\label{4}

Let us start by mentioning the main result:
in undergraduate physics, we learned that when a force $F$ is applied to a spring and stretches it by an amount $x$, in the linear regime these quantities are related by the simple ``Hooke's law'': $F=kx$. 
Recently, in a joint work with Di Gennaro and Good \cite{2108.13435}, we found that in the case of singly rotating black holes in dimensions $d \geqslant 5$, by defining the ``spring constant'' $k$ from the difference of the Hawking temperature between a rotating black hole and a non-rotating one, and by identifying $x$ with the event horizon $r_+$ of the rotating black hole, the maximum value of $F=kx$ in higher dimensions correspond to the instability point at which the black hole fragments into two, almost like an over-stretched spring. Note that black hole fragmentation is entropically allowed in $d \geqslant 5$, unlike in 4-dimensions in which fragmentation is impossible due to Hawking's area theorem. Though there is no fragmentation in $d=4$, the Hookean law still exhibits interesting properties which I will come back to in the following.

Explicitly, the Hawking temperature of a Myers-Perry black hole in $d$-dimensional spacetime is given by
\begin{equation}
	T = \frac{1}{4\pi}\left(\frac{2r_+^{d-4}}{\mu}+ \frac{d-5}{r_+}\right), 
\end{equation}
where $\mu$ is a normalized mass parameter\footnote{Here $\Omega_{d-2}$ is the area of the unit $(d-2)$-dimensional sphere, not to be confused with the angular velocity of the horizon, which I will denote $\Omega_+$.}:
\begin{equation*}
	\mu := \frac{16\pi G}{(d-2)\Omega_{d-2}} M, \qquad \Omega_{d-2} := \frac{2\pi^{\frac{d-1}{2}}}{\Gamma\left(\frac{d-1}{2}\right)}.
\end{equation*}
We can simplify the calculation by assuming an ultra-spinning limit (but this is not necessary) so that $\mu$ can be expressed as $\mu = r_+^{d-5}a^2$, where $a$ is the usual rotation parameter. 
Then the temperature can be re-written as 
\begin{equation}
	T = \frac{1}{2\pi}\left[\frac{d-3}{2 \mu^{\frac{1}{d-3}}} - \underbrace{\left[
	\frac{d-3}{2r_+} \left(\frac{r_+}{a}\right)^{\frac{2}{d-3}}-\frac{d-5}{2r_+} - \frac{r_+}{a^2} \right]}_{=:k}\right].
\end{equation}
When $k=0$ this expression reduces back to the temperature of a static black hole. So in this sense the ``spring constant'' $k$ measures a deviation away from being Schwarzschild. The ``Hookean force'' $F=kr_+$ can be shown to be bounded from above
\begin{equation}
	F = \frac{d-3}{2} \left(\frac{r_+}{a}\right)^{\frac{2}{d-3}}-\frac{d-5}{2} - \frac{r_+^2}{a^2} \lesssim 0.21
\end{equation}
in Planck units. 
We referred to this as a ``force'' only for convenience, it is not really a force, which can be seen by restoring $\hbar, G, c, k_B$ explicitly. Remarkably, despite the naively defined $F=kr_+$, the upper bound corresponds to parameter values at which the black hole becomes unstable and is about to fragment into two Schwarzschild pieces. In terms of the rotation parameter $a$ this happens when $a/r_+ \gtrsim 1.36$ according to Emparan and Myers \cite{0308056}.

This trend holds in $d\geqslant 6$, with the upper bound on $F$ being monotonically decreasing with $d$. The case for $d=4$ is interesting in its own right. There is no fragmentation in this case. However, the expression of $k$ becomes equivalent to $k=M\Omega_+^2$, which is of the same form as the spring constant of an actual spring of mass $m$, namely, $k=m\omega^2$ in undergraduate physics. It is this coincidence that led Good and I to propose the black hole ``spring constant'' in the first place \cite{1412.5432}. This does not hold in higher dimensions, but the definition of $k$ from Hawking temperature \emph{does} extend meaningfully to higher dimensions. In $d=4$ the Hookean law satisfies $F \leqslant 1/4$, with equality attained when the black hole is an extremal Kerr solution. Note that there is no extremal limit in $d\geqslant 6$ singly rotating Myers-Perry spacetimes; for $d=5$ the extremal case is essentially a naked singularity (the horizon coincides with it), but the black hole is likely to fragment before it becomes extremal. 

Despite the fact that $F$ is not a real force, this looks suspiciously like the ``maximum force conjecture'' in GR \cite{0210109,1408.1820,0607090,724159}, which essentially states that the force between two bodies is bounded from above by $1/4$ in its strong form, and by an $O(1)$ value in its weak form \cite{1809.00442}. There are indeed other conjectures concerning the maximum value of physical quantities that can be obtained from the maximum force by multiplying appropriate factors of $\hbar, G, c$ \cite{1803.03271,2105.06650,2109.05973}, although the maximum value is not necessarily $1/4$. Having said that, surprisingly, the value $1/4$ does make some intriguing appearances in our investigations of Myers-Perry black holes, although it is still unclear whether this is just a coincidence. First, as noted previously, the Hookean law satisfies $F \leqslant 1/4$ in all dimensions. More intriguing features are revealed by plotting the sign of the Ruppeiner scalar in the phase space of angular momentum and entropy. 

The Ruppeiner metric \cite{rup} is essentially the Hessian of the entropy function of the black hole parameters,
\begin{equation}\label{ruppeiner}
g^R_{ij}:=-\partial_i\partial_j S(M,N^a),
\end{equation}
interpreted as a Riemannian metric in the phase space.
Here $M$ is the black hole mass, and $N^a$ are other extensive variables of the system (for our case, the angular momentum $J$). {The Ruppeiner scalar, $\mathcal{R}$, is the scalar curvature computed from this metric. In terms of the entropy $S$ and the angular momentum $J$, it is given by
\begin{equation}
\mathcal{R} = -\frac{1}{S} \frac{1-12\cdot\frac{d-5}{d-3}\frac{J^2}{S^2}}{\left(1-4\cdot\frac{d-5}{d-3}\frac{J^2}{S^2}\right)\left(1+4\cdot\frac{d-5}{d-3}\frac{J^2}{S^2}\right)}.
\end{equation}}
If we examine the sign of Ruppeiner scalar in the plot of $S$ against $J$, we will find that \cite{2108.13435} in 6 dimensions and above, there is always a wedge bounded by two straight lines passing through the origin in which the Ruppeiner scalar is positive. The upper line corresponds exactly to $J^2/S^2=1/4$ in $d=6$, along which the Ruppeiner scalar vanishes. As $d$ increases, the wedge region moves with respect to this line, so that $J^2/S^2=1/4$ is contained in the wedge (and thus has positive Ruppeiner scalar). In the large $d$ limit, the line $J^2/S^2=1/4$ tends to the lower boundary line of the wedge. The lower line always corresponds to a diverging Ruppeiner scalar in all dimensions $d \geqslant 6$. 

It has been argued that the Ruppeiner scalar encodes the type of interactions of the underlying degrees of freedom in a statistic mechanical system \cite{0802.1326, 1007.2160}. More precisely, the scalar curvature is positive if the underlying statistical interactions of
the thermodynamical system is repulsive, and likewise it is negative for attractive interactions. 
The system becomes more unstable as the Ruppeiner scalar increases its value \cite{J1,J2}.
A diverging scalar curvature indicates a phase transition. Of course, if the Ruppeiner metric (Eq.(\ref{ruppeiner})) is defined without the negative sign then the interpretation of the interactions would be the opposite. 

Modulo the subtleties of whether this interpretation really applies to black holes because the spacetime metric signature is Lorentzian \cite{1507.06097}, our results suggest that the microstructures of black holes usually have an attractive interaction among themselves. As angular momentum increases, the interaction can turn repulsive and eventually the black hole breaks apart. The fragmentation line is not the same as the phase transition line mentioned above, but is rather always located above the $J^2/S^2=1/4$ line (so the phase transition is not really relevant). As the number of dimension increases, the fragmentation line shifts closer towards the phase transition line. This seems to indicate that as the number of spacetime dimensions increases, it gets harder to break apart the black hole (as the attractive interaction is stronger). We suggested an intuitive picture: imagine the microstructures are like molecules that bond to their neighbors in many directions. With more independent spatial directions, there are more bonds available, and so it becomes increasingly hard to completely break them apart by fast rotation with respect to one particular axis only. Of course this does not tell us what sort of microstructures underlie black hole thermodynamics, but it does perhaps give us a new direction to think about some issues. (Note that the Hookean $F$ should \emph{not} be interpreted as the actual force between the microstructures.) 

It is unlikely that the bound for the Hookean law $F \leqslant 1/4$ is true for all black holes, but this might not be a shortcoming. Rather, it might be an indication that the linear Hookean law $F=kx$ no longer applies. As we suggested in \cite{2108.13435}, it is possible that other black hole solutions (with other matter fields, non-GR etc.) might go beyond the linear spring regime (and the new $F$ might then be bounded by $1/4$, but this is just a guess at this point). After all, for actual springs, how the force scales with the stretched distance $x$ in the non-linear regime depends on what sort of material it is made from. Perhaps black holes would be the same, and the behavior beyond the naively defined Hookean law might reveal more about the different microstructures underlying black holes in different theories. 

\section{Outlook for the Future}\label{5}

In this review I have looked into some aspects of black hole thermodynamics and black hole singularities, as well as their possible connections. 
In particular, I discussed the possibility that some kind of quantum singularities might remain even in QG, that singularities play some roles in guaranteeing a stable ground state exists, as well as
connecting the black hole ``internal energy'' that appears in the first law of black hole thermodynamics with the black hole ADM mass. Since energy is somewhat related to the least action principle, it is also worth mentioning the so-called ``finite-action conjecture'' \cite{331031a0, 1912.12926}, in which the finiteness of the action of a cosmological theory (``total action of the universe'') is tied to the singularities in spacetime. In other words, if a spacetime has no singularity, then the ``action singularities'' (divergence) cannot be avoided and vice-versa. 
Thus the question becomes: is keeping the action finite more important, such that spacetime singularities are relatively small price to pay?
Incidentally, requiring a finite total action also relates to the Weyl curvature hypothesis and the arrow of time issue \cite{1909.01169}, and hence is relevant for gravitational entropy.

I also discussed the ever mysterious microstructures of black holes -- the ``spacetime atoms/molecules'' that form the statistical mechanics basis for black hole thermodynamics, from two perspectives. 
The first is from the point of view of gravitational entropy utilizing the Cartan invariant and the Weyl scalar, which reproduces (up to a constant prefactor) the Bekenstein-Hawking entropy for a spherically symmetric black hole that satisfies the area law. Some aspects of this proposal (as well as some other similar proposals in the literature) remain mysterious, such as why it involves the integral from the singularity to the horizon, regardless of whether $r$ is a spatial or temporal coordinate. As far as the microstructures are concerned, we can conclude that if this proposal is correct, then (1) the Bekenstein-Hawking entropy measures only the gravitational degrees of freedom even when there are some other fields present (e.g., the Maxwell field in the Reissner-Nordstr\"om case); and (2) the microstructures could be different for black holes that do not satisfy the area law, because different combinations of curvature invariants would be required to reproduce their entropy expressions. 

The second perspective is from the point of view of a recently proposed Hookean law of black holes, which also hinted at possibly different microstructures for black holes in different theories of gravity. The Hookean law $F=kx$ for Myers-Perry black holes in GR is bounded by some dimension-dependent value (all less than $1/4$) that also correspond to black hole parameters that, as Emparan and Myers shown \cite{0308056}, the black holes become unstable and should fragment into two Schwarzschild pieces. In this sense it is an interesting and novel aspect of black hole thermodynamics that deserves a deeper study.

I think there are plenty of interesting issues to be further investigated, especially about gravitational entropy, and how this can be properly taken into account by the generalized second law when gravitational waves are emitted during black hole mergers. 
Is the GSL a good guiding principle even in QG? 
The relationship between singularities and black hole thermodynamics should also be looked into on various levels. For example, a recent work by Bousso and Shahbazi-Moghaddam \cite{2201.11132} proved a version of the singularity theorem assuming the existence of ``hyperentropic region'', namely a spacetime region that has an entropy that exceeds the Bekenstein-Hawking entropy of its spatial boundary. Essentially this tells us that a singularity must form if a region has too much entropy. What other fascinating connections are there between singularities and thermodynamics in general?
Can we somehow prescribe the notion of entropy to singularities themselves? (One such attempt in a special case was carried out in \cite{1103.3898}.) This might be needed if we were to understand why two colliding black holes cannot ``annihilate'' each other by converting into pure gravitational waves, by keeping track of the entropy in the gravitational waves and the singularity that might be left behind.
There is also another huge area that has made quite a lot of progress in recent years that I have not reviewed in this work, namely the quantum information perspective, especially insights gained in the contexts of gauge/gravity correspondence (entanglement entropy, complexity, etc.); see \cite{2201.03096,2203.07117,caiqy}. 

On the other hand, I also feel that we need to re-examine some very fundamental issues regarding the nature of Bekenstein-Hawking entropy.
We should perhaps take a step back and ask whether black hole thermodynamics is special in some sense compared to other ``ordinary'' thermodynamical systems.
In the context of black hole evaporation, it is the prevalent point of view among most theoretical physicists (with notable exceptions like Penrose) that quantum information is never lost, and that an evaporating black hole is fundamentally the same as a burning book or a piece of hot coal \cite{1511.01162}, so that the information in a black hole can in principle be recovered from the Hawking radiation. This might or might not be true. Regardless, there is one fundamental difference between black hole and ordinary thermodynamical system: the second law of the latter system is not an absolute law, rather it is a statistical one. There is always a very small probability that some rare configurations of the microstates can occur and so the entropy decreases. To put simply, there are thermal fluctuations that can lower the entropy. In fact, given sufficient time, systems near equilibrium \emph{will} spontaneously fluctuate into lower-entropy states, thus locally reversing the thermodynamic arrow of time (the cosmological implication of such a phenomenon was studied in \cite{1108.0417}). The second law for black holes (Hawking's area law), on the other hand, is absolute. As long as the energy condition is satisfied, the differential geometric proof does not allow the horizon area to ever decrease under classical processes. (Other \emph{dissimilarities} between black holes and ordinary thermodynamical systems were discussed in \cite{Dougherty}.)

Historically, Max Planck once objected Boltzmann's theory of atoms because he believed that the second law should be absolute and not statistical. He thought that the atomic idea was irreconcilably opposed to the law of entropy increase \cite{Helge}. Now, interestingly, as already explained, black hole entropy is ``Planckian'' rather than ``Boltzmannian''. One may wish to argue that the interpretation of black hole area as entropy requires quantum mechanics, or else the entropy would be infinite, since $\hbar$ is in the denominator of the Bekenstein-Hawking formula (though the first law makes sense without quantum mechanics: $\hbar$ cancels out in $T \d S$). So in this sense, Hawking's area theorem is not entirely classical and therefore the energy conditions might not hold absolutely, which in turn allows the area to occasionally decrease. However, usually when we say that the classical energy condition can be violated when quantum effects are considered, we are referring to the \emph{matter sector}. What happens when we consider only the gravitational field? How does the quantum nature of the horizon entropy allow the area to decrease, and how is this tied to the energy condition? 
To some extent, we already understood that the classical null energy condition is a property of gravity that follows from the second law of thermodynamics applied locally to Bekenstein-Hawking entropy associated with patches of null congruences \cite{1511.06460}. But to fully answer this question we may have to first understand the properties of the microstructures of black holes. A possible arena to investigate this issue is via gauge/gravity correspondence, since the Bekenstein-Hawking entropy in the bulk is dual to the entropy of the matter field at the AdS boundary, thus if the matter field entropy is ``Boltzmannian'', so should the Bekenstein-Hawking entropy in some way. On the other hand, we are used to thermodynamics being a manifestation of the underlying statistical mechanics. Is this really true for black holes? Perhaps a different approach in understanding thermodynamics quite independently of statistical mechanics could shed some light on this mystery \cite{2204.04352,9708200,9805005}.

Lastly, let me mention some other quantum aspects of singularities. I have already discussed the possibility that the classical singularities are not completely cured, but rendered more benign in some form in QG. If so, this could be useful for other purposes. For example, it has been suggested that such singularities could induce dimensional reduction effects in QG, which could render QG perturbatively renormalizable \cite{1205.2586,1401.6283}. Penrose, on the other hand, suggested that information is truly lost inside a black hole once it hits the singularity. As a consequence, the relevant phase space volume is reduced \cite{HawkingPenrose}. Imposing suitable boundary conditions at or near the singularity (or whatever replaced it in QG) may also help to resolve the arrow of time problem \cite{0310281,1911.02129,2108.05744}. Curiously, 
there is also the possibility that the arrow of time is nontrivial inside black holes \cite{1911.02129, 0806.3818}. As for cosmic censorship at the quantum level, a curious related result is the discovery of a dynamically generated boundary, dubbed the ``Zeno border'' near the Schwarzschild singularity \cite{1710.07645}. There is no loss of predictability because the singularity cannot be probed by a quantum field. 
This is consistent with the previous results in the literature that Schwarzschild spacetime is quantum complete in the sense that the ground state does not support field configurations near the singularity \cite{1504.05580}; see also \cite{1710.07645,2106.03715}. It is clear that there are many interesting issues to be studied, even after almost half a century of the discovery of black hole thermodynamics.\\

\begin{acknowledgments}
I thank the organizing committee for the invitation to present my research during the SCRI21 meeting, as well as for the opportunity to contribute to the proceeding. It is an honor to participate in such a tribute to Sir Roger Penrose. Instead of focusing on a small aspect that I talked about in the presentation during the meeting, I decided to write a review to organize some of my current thoughts, which span some topics that are related to or inspired by Penrose's own works, including the singularity theorem, cosmic censorship conjecture, and the Weyl curvature hypothesis. I would like to take this opportunity to thank Penrose himself for his remarkable contributions to mathematics, physics, and philosophy, which continue to provide rich ideas for research. I also thank the National Natural Science Foundation of China (No.11922508) for funding support. 
\end{acknowledgments}

\section*{Data Availability}
Data sharing not applicable to this article as no datasets were generated or analyzed during the current study.


\begin{thebibliography}{200}

\bibitem{hawking}
Stephen W. Hawking, ``Black Hole Explosions?'', {\changeurlcolor{vividviolet}\href{https://www.nature.com/articles/248030a0}{Nature \textbf{248} (1974) 30}}.


\bibitem{BCH}
James M. Bardeen, Brandon Carter, Stephen W. Hawking, ``The Four Laws of Black Hole Mechanics'', {\changeurlcolor{vividviolet}\href{https://link.springer.com/article/10.1007/BF01645742}{Commun. Math. Phys. \textbf{31} (1973) 161}}.

\bibitem{b}
Jacob D. Bekenstein, ``Black Holes and Entropy'', {\changeurlcolor{vividviolet}\href{https://journals.aps.org/prd/abstract/10.1103/PhysRevD.7.2333}{Phys. Rev. D \textbf{7} (1973) 2333}}.

\bibitem{1810.01313}
Stephon H. Alexander, Kent Yagi, Nicolas Yunes, ``An Entropy-Area Law for Neutron Stars Near the Black Hole Threshold'', {\changeurlcolor{vividviolet}\href{https://iopscience.iop.org/article/10.1088/1361-6382/aaf14b}{Class. Quant. Grav. \textbf{36} (2019) 1, 015010}}, \href{https://arxiv.org/abs/1810.01313}{[arXiv:1810.01313 [gr-qc]]}.

\bibitem{1910.13198}
Shih-Yuin Lin, ``Seeing Through a Nearly Black Star'', {\changeurlcolor{vividviolet}\href{https://journals.aps.org/prd/abstract/10.1103/PhysRevD.102.025005}{Phys. Rev. D \textbf{102} (2020) 2, 025005}}, \href{https://arxiv.org/abs/1910.13198}{[arXiv:1910.13198 [gr-qc]]}.


\bibitem{area}
Stephen W. Hawking, ``Gravitational Radiation from Colliding Black Holes'', {\changeurlcolor{vividviolet}\href{https://journals.aps.org/prl/abstract/10.1103/PhysRevLett.26.1344}{Phys. Rev. Lett. \textbf{26} (1971) 1344}}.

\bibitem{GSL}
Jacob D. Bekenstein, ``Generalized Second Law of Thermodynamics in Black-Hole Physics'', {\changeurlcolor{vividviolet}\href{https://journals.aps.org/prd/abstract/10.1103/PhysRevD.9.3292}{Phys. Rev. D \textbf{9} (1974) 3292}}.

\bibitem{1710.02724}
David Wallace, ``The Case for Black Hole Thermodynamics, Part I: Phenomenological Thermodynamics'', {\changeurlcolor{vividviolet}\href{https://doi.org/10.1016/j.shpsb.2018.05.002}{Stud. Hist. Phil. Sci. B \textbf{64} (2018) 52}}, \href{https://arxiv.org/abs/1710.02724}{[arXiv:1710.02724 [gr-qc]]}.

\bibitem{1710.02725}
David Wallace, ``The Case for Black Hole Thermodynamics, Part II: Statistical Mechanics'', {\changeurlcolor{vividviolet}\href{https://linkinghub.elsevier.com/retrieve/pii/S1355219818300972}{Stud. Hist. Phil. Sci. B \textbf{66} (2019) 103}}, \href{https://arxiv.org/abs/1710.02725}{[arXiv:1710.02725 [gr-qc]]}.

\bibitem{1903.06276}
Carina E. A. Prunkl, Christopher G. Timpson, ``Black Hole Entropy Is Thermodynamic Entropy'', \href{https://arxiv.org/abs/1903.06276}{[arXiv:1903.06276 [physics.hist-ph]]}.

\bibitem{bousso}
Raphael Bousso, ``The Holographic Principle", 
{\hypersetup{urlcolor=vividviolet}\href{https://journals.aps.org/rmp/abstract/10.1103/RevModPhys.74.825}{Rev. Mod. Phys. \textbf{74}  (2002) 825}}, \href{https://arxiv.org/abs/hep-th/0203101}{[arXiv:hep-th/0203101]}.


\bibitem{1501.00007}
Veronika E. Hubeny, ``The AdS/CFT Correspondence'', {\hypersetup{urlcolor=vividviolet}\href{https://iopscience.iop.org/article/10.1088/0264-9381/32/12/124010}{Class. Quant. Grav. \textbf{32} (2015) 12, 124010}}, \href{https://arxiv.org/abs/1501.00007}{[arXiv:1501.00007 [gr-qc]]}.

\bibitem{1}
Roger Penrose, ``Gravitational Collapse and Space-Time Singularities'', {\hypersetup{urlcolor=vividviolet}\href{https://journals.aps.org/prl/abstract/10.1103/PhysRevLett.14.57}{Phys. Rev. Lett. \textbf{14} (1965) 57}}.

\bibitem{1-2} Roger Penrose, ``Gravitational Collapse: The Role of General Relativity'', Riv. Nuovo Cim. \textbf{1} (1969) 252, {\hypersetup{urlcolor=vividviolet}\href{https://link.springer.com/article/10.1023\%2FA\%3A1016578408204}{Gen. Rel. Grav. \textbf{34} (2002) 1141}}.


\bibitem{9512024}
John F. Donoghue, ``Introduction to the Effective Field Theory Description of Gravity'', \href{https://arxiv.org/abs/gr-qc/9512024}{[arXiv:gr-qc/9512024]}.

\bibitem{0311082}
Cliff P. Burgess, ``Quantum Gravity in Everyday Life: General Relativity as an Effective Field Theory'', {\changeurlcolor{vividviolet}\href{https://link.springer.com/article/10.12942\%2Flrr-2004-5}{Living Rev. Rel. \textbf{7} (2004) 5}}, \href{https://arxiv.org/abs/gr-qc/0311082}{[arXiv:gr-qc/0311082]}.

\bibitem{2204.03869}
James Owen Weatherall, ``Where Does General Relativity Break Down?'', \href{https://arxiv.org/abs/2204.03869}{[arXiv:2204.03869 [physics.hist-ph]]}.

\bibitem{1606.04932}
Sahil Saini, Parampreet Singh, ``Geodesic Completeness and the Lack of Strong Singularities in Effective Loop Quantum Kantowski-Sachs Spacetime'', {\changeurlcolor{vividviolet}\href{https://iopscience.iop.org/article/10.1088/0264-9381/33/24/245019}{Class. Quantum Grav. \textbf{33} (2016) 245019}}, \href{https://arxiv.org/abs/1606.04932}{[arXiv:1606.04932 [gr-qc]]}.

\bibitem{1601.00921}
Stephen W. Hawking, Malcolm J. Perry, Andrew Strominger, ``Soft Hair on Black Holes'', {\changeurlcolor{vividviolet}\href{https://journals.aps.org/prl/abstract/10.1103/PhysRevLett.116.231301}{Phys. Rev. Lett. \textbf{116} (2016) 23, 231301}}, \href{https://arxiv.org/abs/1601.00921}{[arXiv:1601.00921 [hep-th]]}.

\bibitem{1706.07143}
Andrew Strominger, ``Black Hole Information Revisited'', \href{https://arxiv.org/abs/1706.07143}{[arXiv:1706.07143 [hep-th]]}.

\bibitem{1412.8366}
Pisin Chen, Yen Chin Ong, Dong-han Yeom, ``Black Hole Remnants and the Information Loss Paradox'', {\changeurlcolor{vividviolet}\href{https://linkinghub.elsevier.com/retrieve/pii/S0370157315004391}{Phys. Rept. \textbf{603} (2015) 1}}, \href{https://arxiv.org/abs/1412.8366}{[arXiv:1412.8366 [gr-qc]]}.

\bibitem{0601001}
Nima Arkani-Hamed, Lubos Motl, Alberto Nicolis, Cumrun Vafa, ``The String Landscape, Black Holes and Gravity as the Weakest Force'', {\hypersetup{urlcolor=vividviolet}\href{https://iopscience.iop.org/article/10.1088/1126-6708/2007/06/060}{JHEP \textbf{06} (2007) 060}}, \href{https://arxiv.org/abs/hep-th/0601001}{[arXiv:hep-th/0601001]}.

\bibitem{2201.08380}
Daniel Harlow, Ben Heidenreich, Matthew Reece, Tom Rudelius, ``The Weak Gravity Conjecture: A Review'', \href{https://arxiv.org/abs/2201.08380}{[arXiv:2201.08380 [hep-th]]}.

\bibitem{penrose1999}
Roger Penrose, ``The Question of Cosmic Censorship'', {\changeurlcolor{vividviolet}\href{http://adsabs.harvard.edu/full/1999JApA...20..233P}{Jour. of Astrophys. Astronomy \textbf{20} (1999) 233}}.

\bibitem{russo}
Jorge G. Russo, ``On Black Hole Singularities in Quantum Gravity'', {\changeurlcolor{vividviolet}\href{https://www.sciencedirect.com/science/article/abs/pii/0370269394911290}{Phys. Lett. B \textbf{10} (1994) 35}}.


\bibitem{thunder1}
Stephen W. Hawking, John M. Stewart, ``Naked and Thunderbolt Singularities in Black Hole Evaporation'', {\changeurlcolor{vividviolet}\href{https://www.sciencedirect.com/science/article/abs/pii/055032139390410Q?via\%3Dihub}{Nucl. Phys. B \textbf{400} (1993) 393}}, \href{https://arxiv.org/abs/hep-th/9207105}{[arXiv:hep-th/9207105]}.

\bibitem{thunder2}
Stephen W. Hawking, ``Evaporation of Two Dimensional Black Holes'',  In G. Ellis, A. Lanza, \& J. Miller (Eds.), \emph{The Renaissance of General Relativity and Cosmology: A Survey to Celebrate the 65th Birthday of Dennis Sciama}, pp. 274-286, Cambridge, Cambridge University Press, 1993.


\bibitem{0011003}
Shahar Hod, Tsvi Piran, ``Cosmic Censorship: The Role of Quantum Gravity'', {\changeurlcolor{vividviolet}\href{https://link.springer.com/article/10.1023/A:1002098800227}{Gen. Rel. Grav. \textbf{32} (2000) 2333}}, \href{https://arxiv.org/abs/gr-qc/0011003}{[arXiv:gr-qc/0011003]}.

\bibitem{0810.0079}
Shahar Hod, ``Return of the Quantum Cosmic Censor'', {\changeurlcolor{vividviolet}\href{https://www.sciencedirect.com/science/article/pii/S037026930801085X?via\%3Dihub}{Phys. Lett. B \textbf{668} (2008) 346}}, \href{https://arxiv.org/abs/0810.0079}{[arXiv:0810.0079 [gr-qc]]}.

\bibitem{1608.05366}
Marc Casals, Alessandro Fabbri, Cristián Martínez, Jorge Zanelli, ``Quantum Backreaction on Three-Dimensional Black Holes and Naked Singularities'', {\changeurlcolor{vividviolet}\href{https://journals.aps.org/prl/abstract/10.1103/PhysRevLett.118.131102}{Phys. Rev. Lett. \textbf{118} (2017) 131102}}, \href{https://arxiv.org/abs/1608.05366}{[arXiv:1608.05366 [gr-qc]]}.

\bibitem{2109.11968}
Daniele Gregoris, Yen Chin Ong, ``Understanding Gravitational Entropy of Black Holes: A New Proposal via Curvature Invariants'', 	{\changeurlcolor{vividviolet}\href{https://journals.aps.org/prd/abstract/10.1103/PhysRevD.105.104017}{Phys. Rev. D \textbf{105} (2022) 104017}}, \href{https://arxiv.org/abs/2109.11968}{[arXiv:2109.11968 [gr-qc]]}.

\bibitem{2108.13435}
Sofia Di Gennaro, Michael R. R. Good, Yen Chin Ong, ``The Hookean Law of Black Holes and Fragmentation: Insights from Maximum Force Conjecture and Ruppeiner Geometry'', 	{\changeurlcolor{vividviolet}\href{https://journals.aps.org/prresearch/abstract/10.1103/PhysRevResearch.4.023031}{Phys. Rev. Research \textbf{4} (2022) 023031}}, \href{https://arxiv.org/abs/2108.13435}{[arXiv:2108.13435 [gr-qc]]}.

\bibitem{0210109}
Garry W. Gibbons, ``The Maximum Tension Principle in General Relativity'', {\hypersetup{urlcolor=vividviolet}\href{https://link.springer.com/article/10.1023/A:1022370717626}{Found. Phys. \textbf{32} (2002) 1891}}, \href{https://arxiv.org/abs/hep-th/0210109}{[arXiv:hep-th/0210109]}.

\bibitem{1408.1820}
John D. Barrow, G. W. Gibbons, ``Maximum Tension: With and Without a Cosmological Constant'', {\hypersetup{urlcolor=vividviolet}\href{https://academic.oup.com/mnras/article/446/4/3874/2892906}{Mon. Not. Roy. Astron. Soc. \textbf{446} (2014) 3874}}, \href{https://arxiv.org/abs/1408.1820}{[arXiv:1408.1820 [gr-qc]]}.

\bibitem{0607090}
Christoph Schiller, ``General Relativity and Cosmology Derived From Principle of Maximum Power or Force'',  {\hypersetup{urlcolor=vividviolet}\href{https://link.springer.com/article/10.1007\%2Fs10773-005-4835-2}{Int. Jour. Theo. Phys. \textbf{44} (2005) 1629}}, \href{https://arxiv.org/abs/physics/0607090}{[arXiv:physics/0607090 [physics.gen-ph]]}.

\bibitem{724159}
Christoph Schiller, ``Simple Derivation of Minimum Length, Minimum Dipole Moment and Lack of Space-Time Continuity'', {\hypersetup{urlcolor=vividviolet}\href{https://link.springer.com/article/10.1007\%2Fs10773-005-9018-7}{Int. J. Theor. Phys. \textbf{45} (2006) 221}}.


\bibitem{2005.07032}
Yen Chin Ong, ``Spacetime Singularities and Cosmic Censorship Conjecture: A Review with Some Thoughts'',  {\changeurlcolor{vividviolet}\href{https://www.worldscientific.com/doi/abs/10.1142/S0217751X20300070}{Int. J. Mod. Phys. A \textbf{35} (2020) 14, 14}}, \href{https://arxiv.org/abs/2005.07032}{[arXiv:2005.07032 [gr-qc]]}.

\bibitem{2112.08531}
Karen Crowther, Sebastian De Haro, ``Four Attitudes Towards Singularities in the Search for a Theory of Quantum Gravity'', \href{https://arxiv.org/abs/2112.08531}{[arXiv:2112.08531 [gr-qc]]}.

\bibitem{1903.11704}
Christopher S. Reynolds, ``Observing Black Holes Spin'', {\changeurlcolor{vividviolet}\href{https://www.nature.com/articles/s41550-018-0665-z}{Nature Astron. \textbf{3} (2019) 1, 41}}, \href{https://arxiv.org/abs/1903.11704}{[arXiv:1903.11704 [astro-ph.HE]]}.

\bibitem{jose}
Jos\'e M. M. Senovilla, David Garfinkle, ``The 1965 Penrose Singularity Theorem'', {\changeurlcolor{vividviolet}\href{https://iopscience.iop.org/article/10.1088/0264-9381/32/12/124008}{Class. Quant. Grav. \textbf{32} (2015) 124008}}, \href{https://arxiv.org/abs/1410.5226}{[arXiv:1410.5226 [gr-qc]]}.

\bibitem{1012.6038}
Christopher J. Fewster, Gregory J. Galloway, ``Singularity Theorems From Weakened Energy Conditions'', {\changeurlcolor{vividviolet}\href{https://iopscience.iop.org/article/10.1088/0264-9381/28/12/125009}{Class. Quantum Grav. \textbf{28} (2011) 125009}}, \href{https://arxiv.org/abs/1012.6038}{[arXiv:1012.6038 [gr-qc]]}.

\bibitem{1907.13604}
Christopher J. Fewster, Eleni-Alexandra Kontou, ``A New Derivation of Singularity Theorems With Weakened Energy Hypotheses'', {\changeurlcolor{vividviolet}\href{https://iopscience.iop.org/article/10.1088/1361-6382/ab685b}{Class. Quant. Grav. \textbf{37} (2020) 6, 065010}}, \href{https://arxiv.org/abs/1907.13604}{[arXiv:1907.13604 [gr-qc]]}.

\bibitem{2012.11569}
Ben Freivogel, Eleni-Alexandra Kontou, Dimitrios Krommydas, ``The Return of the Singularities: Applications of the Smeared Null Energy Condition'', {\changeurlcolor{vividviolet}\href{https://scipost.org/10.21468/SciPostPhys.13.1.001}{SciPost Phys. \textbf{13} (2022) 001}}, \href{https://arxiv.org/abs/2012.11569}{[arXiv:2012.11569 [gr-qc]]}.

\bibitem{2108.12668}
Christopher J. Fewster, Eleni-Alexandra Kontou, ``A Semiclassical Singularity Theorem'', {\changeurlcolor{vividviolet}\href{https://iopscience.iop.org/article/10.1088/1361-6382/ac566b}{Class. Quant. Grav. \textbf{39} (2022) 7, 075028}}, \href{https://arxiv.org/abs/2108.12668}{[arXiv:2108.12668 [gr-qc]]}.

\bibitem{1909.07348}
Ettore Minguzzi, ``A Gravitational Collapse Singularity Theorem Consistent With Black Hole Evaporation'', {\changeurlcolor{vividviolet}\href{https://link.springer.com/article/10.1007/s11005-020-01295-9}{Lett. Math. Phys. \textbf{110} (2020) 2383}}, \href{https://arxiv.org/abs/1909.07348}{[arXiv:1909.07348 [gr-qc]]}.

\bibitem{0603158}
Sergei L. Dubovsky, Sergey M. Sibiryakov, ``Spontaneous Breaking of Lorentz Invariance, Black Holes and Perpetuum Mobile of the 2nd Kind'', {\changeurlcolor{vividviolet}\href{https://linkinghub.elsevier.com/retrieve/pii/S0370269306006757}{Phys. Lett. B \textbf{638} (2006) 509}}, \href{https://arxiv.org/abs/hep-th/0603158}{[arXiv:hep-th/0603158]}.

\bibitem{0702124}
Christopher Eling, Brendan Z. Foster, Ted Jacobson, Aron C. Wall, ``Lorentz Violation and Perpetual Motion'', {\changeurlcolor{vividviolet}\href{https://journals.aps.org/prd/abstract/10.1103/PhysRevD.75.101502}{Phys. Rev. D \textbf{75} (2007) 101502}}, \href{https://arxiv.org/abs/hep-th/0702124}{[arXiv:hep-th/0702124]}.

\bibitem{0811.0943}
Gerold Betschart, E. Kant, Frans R. Klinkhamer, ``Lorentz Violation and Black-Hole Thermodynamics'', {\changeurlcolor{vividviolet}\href{https://www.sciencedirect.com/science/article/abs/pii/S0550321309000984?via\%3Dihub}{Nucl. Phys. B \textbf{815} (2009) 198}}, \href{https://arxiv.org/abs/0811.0943}{[arXiv:0811.0943 [hep-th]]}.

\bibitem{0804.2720}
Ted Jacobson, Aron C. Wall, ``Black Hole Thermodynamics and Lorentz Symmetry'', {\changeurlcolor{vividviolet}\href{https://link.springer.com/article/10.1007/s10701-010-9423-5}{Found. Phys. \textbf{40} (2010) 1076}}, \href{https://arxiv.org/abs/0804.2720}{[arXiv:0804.2720 [hep-th]]}.


\bibitem{eddington}
Arthur Eddington, ``New Pathways in Science'', \emph{Messenger Lectures}, Cambridge University Press, Cambridge, 1935.

\bibitem{Price3}
Huw Price,
``Time’s Arrow and Eddington’s Challenge'', {\hypersetup{urlcolor=vividviolet}\href{https://link.springer.com/chapter/10.1007\%2F978-3-0348-0359-5_6}{Prog. Math. Phys. \textbf{63} (2013) 187}}.

\bibitem{1010.5513}
Aron C. Wall, ``The Generalized Second Law implies a Quantum Singularity Theorem'', {\hypersetup{urlcolor=vividviolet}\href{https://iopscience.iop.org/article/10.1088/0264-9381/30/16/165003}{Class. Quant. Grav. \textbf{30} (2013) 165003}}, Class. Quant. Grav. \textbf{30} (2013) 199501 (erratum), \href{https://arxiv.org/abs/1010.5513}{[arXiv:1010.5513 [gr-qc]]}.

\bibitem{ghs}
David Garfinkle, Gary T. Horowitz, Andrew Strominger, ``Charged Black Holes in String Theory'', {\hypersetup{urlcolor=vividviolet}\href{https://journals.aps.org/prd/abstract/10.1103/PhysRevD.43.3140}{Phys. Rev. D \textbf{43} (1991) 3140}} [{\hypersetup{urlcolor=vividviolet}\href{https://journals.aps.org/prd/abstract/10.1103/PhysRevD.45.3888}{Erratum ibid. D \textbf{45} (1992) 3888}}].

\bibitem{g}
Gary Gibbons, ``Antigravitating Black Hole Solitons with Scalar Hair in $\mathcal{N} = 4$ Supergravity'',
{\hypersetup{urlcolor=vividviolet}\href{https://linkinghub.elsevier.com/retrieve/pii/0550321382901705}{Nucl. Phys. B \textbf{207} (1982) 337}}.

\bibitem{gm}
 Gary Gibbons, Kei-ichi Maeda,``Black Holes and Membranes in Higher Dimensional Theories with Dilaton Fields'', {\hypersetup{urlcolor=vividviolet}\href{https://www.sciencedirect.com/science/article/pii/0550321388900065?via\%3Dihub}{Nucl. Phys. B \textbf{298} (1988) 741}}.

\bibitem{HW}
William A. Hiscock, Lance D. Weems, ``Evolution of Charged Evaporating Black Holes'', {\hypersetup{urlcolor=vividviolet}\href{http://refhub.elsevier.com/S0370-1573(15)00439-1/sbref222}{Phys. Rev. D \textbf{41} (1990) 1142}}.

\bibitem{gibbons}
Gary W. Gibbons, ``Vacuum Polarization and the Spontaneous Loss of Charge by Black Holes'', {\hypersetup{urlcolor=vividviolet}\href{https://projecteuclid.org/euclid.cmp/1103899346}{Comm. Math. Phys. \textbf{44} (1975) 245}}.

\bibitem{2102.05519}
Shahar Hod, ``Hawking Radiation May Violate the Penrose Cosmic Censorship Conjecture'', {\hypersetup{urlcolor=vividviolet}\href{https://www.worldscientific.com/doi/abs/10.1142/S0218271819440231}{Int. Jour. Mod. Phys. D \textbf{28} (2019) 1944023}}, \href{https://arxiv.org/abs/2102.05519v1}{[arXiv:2102.05519 [gr-qc]]}.


\bibitem{1907.07490}
Yen Chin Ong, Yuan Yao, ``Charged Particle Production Rate from Cosmic Censorship in Dilaton Black Hole Spacetimes'', {\hypersetup{urlcolor=vividviolet}\href{https://link.springer.com/article/10.1007\%2FJHEP10\%282019\%29129}{JHEP \textbf{10} (2019) 129}}, \href{https://arxiv.org/abs/1907.07490}{[arXiv:1907.07490 [gr-qc]]}.

\bibitem{1305.2564}
Kiyoshi Shiraishi, ``Superradiance From a Charged Dilaton Black Hole'', {\hypersetup{urlcolor=vividviolet}\href{https://www.worldscientific.com/doi/abs/10.1142/S0217732392002858}{Mod. Phys. Lett. A \textbf{7} (1992) 3449}}, \href{https://arxiv.org/abs/1305.2564v1}{[arXiv:1305.2564 [hep-th]]}.

\bibitem{2103.05516}
Sofia Di Gennaro, Yen Chin Ong, ``How Not to Extract Information From Black Holes: Cosmic Censorship as a Guiding Principle'', {\hypersetup{urlcolor=vividviolet}\href{https://www.sciencedirect.com/science/article/pii/S0370269322002465?via\%3Dihub}{Phys. Lett. B \textbf{829} (2022) 137112}}, \href{https://arxiv.org/abs/2103.05516}{[arXiv:2103.05516 [hep-th]]}.

\bibitem{2103.00257}
Chen-Hao Wu, Ya-Peng Hu, Hao Xu, ``Hawking Evaporation of Einstein-Gauss-Bonnet Ads Black Holes in $D \geqslant$ Dimensions'', {\hypersetup{urlcolor=vividviolet}\href{https://link.springer.com/article/10.1140/epjc/s10052-021-09140-6}{Eur. Phys. J. C \textbf{81} (2021) 4, 351}}, \href{https://arxiv.org/abs/2103.00257}{[arXiv:2103.00257 [hep-th]]}.


\bibitem{2205.13006}
Fabrizio Corelli, Marina De Amicis, Taishi Ikeda, Paolo Pani, ``What Is the Fate of Hawking Evaporation in Gravity Theories With Higher Curvature Terms?'', \href{https://arxiv.org/abs/2205.13006}{[arXiv:2205.13006 [gr-qc]]}.


\bibitem{2205.13007}
Fabrizio Corelli, Marina De Amicis, Taishi Ikeda, Paolo Pani, ``Nonperturbative Gedanken Experiments in Einstein-Dilaton-Gauss-Bonnet Gravity: Nonlinear Transitions and Tests of the Cosmic Censorship Beyond General Relativity'', \href{https://arxiv.org/abs/2205.13007}{[arXiv:2205.13007 [gr-qc]]}.


\bibitem{2207.10692}
Pedro G. S. Fernandes, David J. Mulryne, Jorge F. M. Delgado, ``Exploring the Small Mass Limit of Stationary Black Holes in Theories with Gauss-Bonnet Terms'', \href{https://arxiv.org/abs/2207.10692}{[arXiv:2207.10692 [gr-qc]]}.

\bibitem{9503062}
Gary T. Horowitz, Robert Myers, ``The Value of Singularities'', 	{\hypersetup{urlcolor=vividviolet}\href{https://link.springer.com/article/10.1007\%2FBF02113073}{Gen. Rel. Grav. \textbf{27} (1995) 915}}, \href{https://arxiv.org/abs/gr-qc/9503062}{[arXiv:gr-qc/9503062]}.

\bibitem{AD}
Larry F. Abbott, Stanley Deser, ``Stability of Gravity with a Cosmological Constant'',
{\hypersetup{urlcolor=vividviolet}\href{https://www.sciencedirect.com/science/article/abs/pii/0550321382900499}{Nucl. Phys. B \textbf{195} (1982) 76}}.

\bibitem{0506057}
Stanley Deser, Inanc Kanik, Bayram Tekin, ``Conserved Charges of Higher D Kerr-AdS Spacetimes'', {\hypersetup{urlcolor=vividviolet}\href{https://iopscience.iop.org/article/10.1088/0264-9381/22/17/001}{Class. Quant. Grav. \textbf{22} (2005) 3383}}, \href{https://arxiv.org/abs/gr-qc/0506057}{[arXiv:gr-qc/0506057]}.

\bibitem{1506.01248}
Brett McInnes, Yen Chin Ong, ``A Note on Physical Mass and the Thermodynamics of AdS-Kerr Black Holes'', 	{\hypersetup{urlcolor=vividviolet}\href{https://iopscience.iop.org/article/10.1088/1475-7516/2015/11/004}{JCAP \textbf{11} (2015) 004}}, \href{https://arxiv.org/abs/1506.01248}{[arXiv:1506.01248 [gr-qc]]}.

\bibitem{1411.0833}
Meng-Sen Ma, Ren Zhao, ``Corrected Form of the First Law of Thermodynamics for Regular Black Holes'', {\hypersetup{urlcolor=vividviolet}\href{https://iopscience.iop.org/article/10.1088/0264-9381/31/24/245014}{Class. Quant. Grav. \textbf{31} (2014) 245014}}, \href{https://arxiv.org/abs/1411.0833}{[arXiv:1411.0833 [gr-qc]]}.

\bibitem{2103.14413}
Chen Lan, Yan-Gang Miao, ``Gliner Vacuum, Self-consistent Theory of Ruppeiner Geometry for Regular Black Holes'', \href{https://arxiv.org/abs/2103.14413}{[arXiv:2103.14413 [gr-qc]]}.

\bibitem{2208.12713}
Ram\'on Torres, ``Regular Rotating Black Holes: A Review'', \href{https://arxiv.org/abs/2208.12713}{[arXiv:2208.12713 [gr-qc]]}.

\bibitem{2209.10612}
Alfio Bonanno, Amir-Pouyan Khosravi, Frank Saueressig, ``Regular Evaporating Black Holes With Stable Cores'', \href{https://arxiv.org/abs/2209.10612}{[arXiv:2209.10612 [gr-qc]]}.

\bibitem{1805.02675}
Ra\'{u}l Carballo-Rubio, Francesco Di Filippo, Stefano Liberati, Costantino Pacilio, Matt Visser, ``On the Viability of Regular Black Holes'', {\hypersetup{urlcolor=vividviolet}\href{https://link.springer.com/article/10.1007/JHEP07(2018)023}{JHEP \textbf{2018} (2018) 23}}, \href{https://arxiv.org/abs/1805.02675}{[arXiv:1805.02675 [gr-qc]]}.

\bibitem{2101.05006}
Ra\'{u}l Carballo-Rubio, Francesco Di Filippo, Stefano Liberati, Costantino Pacilio, Matt Visser, ``Inner Horizon Instability and the Unstable Cores of Regular Black Holes'', {\hypersetup{urlcolor=vividviolet}\href{https://link.springer.com/article/10.1007/JHEP05(2021)132}{JHEP \textbf{05} (2021) 132}}, \href{https://arxiv.org/abs/2101.05006}{[arXiv:2101.05006 [gr-qc]]}.

\bibitem{2205.13556}
Raúl Carballo-Rubio, Francesco Di Filippo, Stefano Liberati, Costantino Pacilio, Matt Visser, ``Regular Black Holes Without Mass Inflation Instability'', \href{https://arxiv.org/abs/2205.13556}{[arXiv:2205.13556 [gr-qc]]}.

\bibitem{2207.08864}
Edgardo Franzin, Stefano Liberati, Jacopo Mazza, Vania Vellucci, ``Stable Rotating Regular Black Holes'', \href{https://arxiv.org/abs/2207.08864}{[arXiv:2207.08864 [gr-qc]]}.

\bibitem{1311.4970}
Jonathan Luk, ``Weak Null Singularities in General Relativity'',  {\hypersetup{urlcolor=vividviolet}\href{https://www.ams.org/journals/jams/2018-31-01/S0894-0347-2017-00888-9/}{J. Am. Math. Soc. \textbf{31} (2018) 1}}, \href{https://arxiv.org/abs/1311.4970}{[arXiv:1311.4970 [gr-qc]]}.

\bibitem{1710.01722}
Mihalis Dafermos, Jonathan Luk, ``The Interior of Dynamical Vacuum Black Holes I: The $C^0$-Stability of the Kerr Cauchy Horizon'', \href{https://arxiv.org/abs/1710.01722}{[arXiv:1710.01722 [gr-qc]]}.

\bibitem{1704.05790}
Maxime Van de Moortel, ``Stability and Instability of the Sub-Extremal Reissner-Nordstr\"om Black Hole Interior for the Einstein-Maxwell-Klein-Gordon Equations in Spherical Symmetry'', {\hypersetup{urlcolor=vividviolet}\href{https://link.springer.com/article/10.1007/s00220-017-3079-3}{Commun. Math. Phys. \textbf{360}  (2018) 103}}, \href{https://arxiv.org/abs/1704.05790}{[arXiv:1704.05790 [gr-qc]]}.


\bibitem{1912.10890}
Maxime Van de Moortel, ``The Breakdown of Weak Null Singularities Inside Black Holes'', \href{https://arxiv.org/abs/1912.10890}{[arXiv:1912.10890 [gr-qc]]}.


\bibitem{2001.11156}
Maxime Van de Moortel, ``Mass Inflation and the $C^2$-Inextendibility of Spherically Symmetric Charged Scalar Field Dynamical Black Holes'', {\hypersetup{urlcolor=vividviolet}\href{https://link.springer.com/article/10.1007/s00220-020-03923-w}{Commun. Math. Phys. \textbf{382} (2021) 2, 1263}}, \href{https://arxiv.org/abs/2001.11156}{[arXiv:2001.11156 [gr-qc]]}.

\bibitem{2105.04604}
Christoph Kehle, Maxime Van de Moortel, ``Strong Cosmic Censorship in the Presence of Matter: The Decisive Effect of Horizon Oscillations on the Black Hole Interior Geometry'', \href{https://arxiv.org/abs/2105.04604}{[arXiv:2105.04604 [gr-qc]]}.

\bibitem{1005.2999}
Donald Marolf, ``The Dangers of Extremes'', {\hypersetup{urlcolor=vividviolet}\href{https://link.springer.com/article/10.1007/s10714-010-1027-z}{Gen. Rel. Grav. \textbf{42} (2010) 2337}}, \href{https://arxiv.org/abs/1005.2999}{[arXiv:1005.2999 [gr-qc]]}.

\bibitem{2206.07001}
Raphael Bousso, Arvin Shahbazi-Moghaddam, ``Quantum Singularities'', \href{https://arxiv.org/abs/2206.07001}{[arXiv:2206.07001 [hep-th]]}.

\bibitem{2011.01004}
Matan Grinberg, Juan Maldacena, ``Proper Time to the Black Hole Singularity From Thermal One-Point Functions'', {\hypersetup{urlcolor=vividviolet}\href{https://link.springer.com/article/10.1007/JHEP03(2021)131}{JHEP \textbf{03} (2021) 131}}, \href{https://arxiv.org/abs/2011.01004}{[arXiv:2011.01004 [hep-th]]}.

\bibitem{2206.00198}
Brett McInnes, ``Inside Flat Event Horizons'', \href{https://arxiv.org/abs/2206.00198}{[arXiv:2206.00198 [gr-qc]]}.

\bibitem{penrosetime}
Roger Penrose, ``Singularities and Time-Asymmetry'', pp. 581-638 in \emph{General Relativity:
An Einstein Centenary Survey}, eds. S. W. Hawking, W. Israel, Cambridge University Press, Cambridge, 1979.

\bibitem{123}
Roger Penrose, ``Before the Big Bang: An Outrageous New Perspective and Its Implications for Particle Physics'', Contribution to: 10th European Particle Accelerator Conference (EPAC 06), {\hypersetup{urlcolor=vividviolet}\href{https://accelconf.web.cern.ch/e06/PAPERS/THESPA01.PDF}{Conf. Proc. C \textbf{060626} (2006) 2759}}.

\bibitem{0909.3983}
Chas A. Egan, Charles H. Lineweaver, ``A Larger Estimate of the Entropy of the Universe'', 	{\hypersetup{urlcolor=vividviolet}\href{https://iopscience.iop.org/article/10.1088/0004-637X/710/2/1825}{Astrophys. J. \textbf{710} (2010) 1825}}, \href{https://arxiv.org/abs/0909.3983}{[arXiv:0909.3983 [astro-ph.CO]]}.

\bibitem{0711.1656}
Brett McInnes, ``The Arrow Of Time In The Landscape'', \href{https://arxiv.org/abs/0711.1656}{[arXiv:0711.1656 [hep-th]]}.

\bibitem{0609095}
Gary W. Gibbons, Neil Turok, ``The Measure Problem in Cosmology'', 	{\hypersetup{urlcolor=vividviolet}\href{https://journals.aps.org/prd/abstract/10.1103/PhysRevD.77.063516}{Phys. Rev. D \textbf{77} (2008) 063516}}, \href{https://arxiv.org/abs/hep-th/0609095}{[arXiv:hep-th/0609095]}.

\bibitem{wcc1}
Roger Penrose, ``Singularities and Time-Asymmetry", in { \it General Relativity, An Einstein Centenary Survey}, (eds. Stephen William Hawking and Werner Israel, Cambridge University Press, Cambridge, 1979).

\bibitem{1711.06480}
Martin Lesourd, ``Hawking's Area Theorem With a Weaker Energy Condition'', {\hypersetup{urlcolor=vividviolet}\href{https://link.springer.com/article/10.1007/s10714-018-2377-1}{Gen. Rel. Grav. \textbf{50} (2018) 6, 61}}, \href{https://arxiv.org/abs/1711.06480}{[arXiv:1711.06480 [gr-qc]]}.

\bibitem{2012.04486}
Maximiliano Isi, Will M. Farr, Matthew Giesler, Mark A. Scheel, Saul A. Teukolsky, ``Testing the Black-Hole Area Law With GW150914'', {\hypersetup{urlcolor=vividviolet}\href{https://journals.aps.org/prl/abstract/10.1103/PhysRevLett.127.011103}{Phys. Rev. Lett. \textbf{127} (2021) 011103}}, \href{https://arxiv.org/abs/2012.04486}{[arXiv:2012.04486 [gr-qc]]}.


\bibitem{1811.12283}
Ram Brustein, Allan J.M. Medved, Kent Yagi, ``Lower Limit on the Entropy of Black Holes as Inferred From Gravitational Wave Observations'', {\hypersetup{urlcolor=vividviolet}\href{https://journals.aps.org/prd/abstract/10.1103/PhysRevD.100.104009}{Phys. Rev. D \textbf{100} (2019) 10, 104009}}, \href{https://arxiv.org/abs/1811.12283}{[arXiv:1811.12283 [gr-qc]]}.


\bibitem{smolin}
Lee Smolin, ``On the Intrinsic Entropy of the Gravitational Field'', {\hypersetup{urlcolor=vividviolet}\href{https://link.springer.com/article/10.1007/BF00761902}{Gen. Rel. Grav. \textbf{17} (1985) 417}}.

\bibitem{1303.5612}
Timothy Clifton, George F R Ellis, Reza Tavakol, ``A Gravitational Entropy Proposal'', {\hypersetup{urlcolor=vividviolet}\href{https://iopscience.iop.org/article/10.1088/0264-9381/30/12/125009}{Class .Quant. Grav. \textbf{30} (2013) 125009}}, \href{https://arxiv.org/abs/1303.5612}{[arXiv:1303.5612 [gr-qc]]}.

\bibitem{2205.02985}
Fernando A. Piza\~{n}a, Roberto A. Sussman, Juan Carlos Hidalgo, ``Gravitational Entropy in Szekeres Class I Models'', {\hypersetup{urlcolor=vividviolet}\href{https://iopscience.iop.org/article/10.1088/1361-6382/ac851a}{Class. Quantum Grav. \textbf{39} (2022) 185005}}, \href{https://arxiv.org/abs/2205.02985}{[arXiv:2205.02985 [gr-qc]]}.

\bibitem{2004.10222}
Daniele Gregoris, Yen Chin Ong, Bin Wang, ``Thermodynamics of Shearing Massless Scalar Field Spacetimes is Inconsistent With the Weyl Curvature Hypothesis'', {\hypersetup{urlcolor=vividviolet}\href{https://journals.aps.org/prd/abstract/10.1103/PhysRevD.102.023539}{Phys. Rev. D \textbf{102} (2020) 023539}}, \href{https://arxiv.org/abs/2004.10222}{[arXiv:2004.10222 [gr-qc]]}.

\bibitem{Wald}
Robert M. Wald, \emph{General Relativity}, University of Chicago Press, 1984.

\bibitem{2201.01939}
Brett McInnes, ``Planar Black Holes as a Route to Understanding the Weak Gravity Conjecture'', {\hypersetup{urlcolor=vividviolet}\href{https://www.sciencedirect.com/science/article/pii/S055032132200284X?via\%3Dihub}{Nucl. Phys. B \textbf{983} (2022) 115933}}, \href{https://arxiv.org/abs/2201.01939}{[arXiv:2201.01939 [gr-qc]]}.

\bibitem{1107.5821}
Ruth Gregory, ``The Gregory-Laflamme Instability'', In G. Horowitz (Ed.), \emph{Black Holes in Higher Dimensions}, pp. 29-43, Cambridge: Cambridge University Press, 2012, \href{https://arxiv.org/abs/1107.5821}{[arXiv:1107.5821 [gr-qc]]}.

\bibitem{1812.05017}
Tomas Andrade, Roberto Emparan, David Licht, Raimon Luna, ``Cosmic Censorship Violation in Black Hole Collisions in Higher Dimensions'', {\hypersetup{urlcolor=vividviolet}\href{https://link.springer.com/article/10.1007\%2FJHEP04\%282019\%29121}{JHEP \textbf{04} (2019) 121}}, \href{https://arxiv.org/abs/1812.05017}{[arXiv:1812.05017 [hep-th]]}.

\bibitem{9601029}
Andrew Strominger, Cumrun Vafa, ``Microscopic Origin of the Bekenstein-Hawking Entropy'', {\hypersetup{urlcolor=vividviolet}\href{https://linkinghub.elsevier.com/retrieve/pii/0370269396003450}{Phys. Lett. B \textbf{379} (1996) 99}},
\href{https://arxiv.org/abs/hep-th/9601029}{[arXiv:hep-th/9601029]}.

\bibitem{hara}
Tetsuya Hara, Keita Sakai, Shuhei Kunitomo, Daigo Kajiura, ``The Entropy Increase During the Black Hole Formation'', in \emph{Black Holes from Stars to Galaxies -- Across the Range of Masses}, Ed. V. Karas and G. Matt., Proceedings of IAU Symposium 238, UK: Cambridge University Press, 2007, pp.{\hypersetup{urlcolor=vividviolet}\href{http://adsabs.harvard.edu/full/2007IAUS..238..377H}{377}}.

\bibitem{2206.11870}
Erik Aurell, ``The Double Doors of the Horizon'', \href{https://arxiv.org/abs/2206.11870v2}{[arXiv:2206.11870 [gr-qc]]}.


\bibitem{1611.04044}
Sina Bahrami, ``Saturating The Bekenstein-Hawking Entropy Bound With Initial Data Sets For Gravitational Collapse'', {\hypersetup{urlcolor=vividviolet}\href{https://journals.aps.org/prd/abstract/10.1103/PhysRevD.95.026006}{Phys. Rev. D \textbf{95} (2017) 026006}}, \href{https://arxiv.org/abs/1611.04044}{[arXiv:1611.04044 [gr-qc]]}.

\bibitem{epjcbh1}
Nan Li, Xiao-Long Li, and Shu-Peng Song,   ``An Exploration of the Black Hole Entropy via the Weyl Tensor",
{\hypersetup{urlcolor=vividviolet}\href{https://link.springer.com/article/10.1140/epjc/s10052-016-3960-9}{Eur. Phys. Jour. C \textbf{76} (2016) 111}}, \href{https://arxiv.org/abs/1510.09027}{[arXiv:gr-qc/1510.09027]}.

\bibitem{cartan}
Dario Brooks, Paul-Christopher Chavy-Waddy, Alan Albert Coley, Adam Forget, Daniele Gregoris, Malcolm A.H.
MacCallum, David Duncan McNutt,  ``Cartan Invariants and Event Horizon Detection",
{\hypersetup{urlcolor=vividviolet}\href{https://link.springer.com/article/10.1007/s10714-018-2358-4}{ Gen. Rel. Grav.  \textbf{50} (2018)  37}}, \href{https://arxiv.org/abs/1709.03362}{[arXiv:gr-qc/1709.03362]}; ``Correction to: Cartan invariants and event horizon detection",
{\hypersetup{urlcolor=vividviolet}\href{https://link.springer.com/article/10.1007/s10714-020-2659-2}{ Gen. Rel. Grav.  \textbf{52} (2020)  6}}.


\bibitem{1010.5844}
Ariel Edery, Benjamin Constantineau, ``Extremal Black Holes, Gravitational Entropy and Nonstationary Metric Fields'', {\hypersetup{urlcolor=vividviolet}\href{https://iopscience.iop.org/article/10.1088/0264-9381/28/4/045003}{Class. Quant. Grav. \textbf{28} (2011) 045003}}, \href{https://arxiv.org/abs/1010.5844}{[arXiv:gr-qc/1010.5844]}.


\bibitem{entropyf1}
Iver Brevik, Shin'ichi Nojiri, Sergei D. Odintsov, and Luciano Vanzo,  ``Entropy and Universality of the Cardy-Verlinde Formula in a Dark Energy Universe",
{\hypersetup{urlcolor=vividviolet}\href{https://journals.aps.org/prd/abstract/10.1103/PhysRevD.70.043520}{ Phys. Rev. D   \textbf{70} (2004)  043520}}, \href{https://arxiv.org/abs/hep-th/0401073}{[arXiv:hep-th/0401073]}.

\bibitem{entropyf2}
Guido Cognola, Emilio Elizalde, Shin'ichi Nojiri, Sergei D. Odintsov, Sergio Zerbini,  ``One-Loop $f(R)$ Gravity in de Sitter Universe",
{\hypersetup{urlcolor=vividviolet}\href{https://iopscience.iop.org/article/10.1088/1475-7516/2005/02/010}{ JCAP   \textbf{02} (2005)  010}}, \href{https://arxiv.org/abs/hep-th/0501096}{[arXiv:hep-th/0501096]}.

\bibitem{entropyf3}
M. Akbar, Rong-Gen Cai,  ``Friedmann Equations of FRW Universe in Scalar-Tensor Gravity, $F(R)$ Gravity and First Law of Thermodynamics",
{\hypersetup{urlcolor=vividviolet}\href{https://www.sciencedirect.com/science/article/pii/S0370269306002255?via\%3Dihub}{ Phys. Lett. B \textbf{635} (2006)  7}}, \href{https://arxiv.org/abs/hep-th/0602156}{[arXiv:hep-th/0602156]}.
		
\bibitem{entropyf4}
Yungui Gong, Anzhong Wang,  ``Friedmann Equations and Thermodynamics of Apparent Horizons",
{\hypersetup{urlcolor=vividviolet}\href{https://journals.aps.org/prl/abstract/10.1103/PhysRevLett.99.211301}{Phys. Rev. Lett. \textbf{99} (2007)  211301}}, \href{https://arxiv.org/abs/0704.0793}{[arXiv:hep-th/0704.0793]}.
		
\bibitem{entropyf5}
Ram Brustein, Dan Gorbonos, Merav Hadad,  ``Wald's Entropy Is Equal to a Quarter of the Horizon Area in Units of the Effective Gravitational Coupling",
{\hypersetup{urlcolor=vividviolet}\href{https://journals.aps.org/prd/abstract/10.1103/PhysRevD.79.044025}{Phys. Rev. D  \textbf{79} (2009)  044025}}, \href{https://arxiv.org/abs/0712.3206}{[arXiv:hep-th/0712.3206]}.

\bibitem{emre}
Emre Dil, ``Gravitational Entropy of a Schwarzschild-Type Black Hole'', {\hypersetup{urlcolor=vividviolet}\href{https://link.springer.com/article/10.3938/jkps.69.6}{Jour. Korean Phys. Soc. \textbf{69} (2016) 6}}.

\bibitem{0311240}
Rong-Gen Cai, ``A Note on Thermodynamics of Black Holes in Lovelock Gravity'', 	{\hypersetup{urlcolor=vividviolet}\href{https://www.sciencedirect.com/science/article/abs/pii/S0370269304000966?via\%3Dihub}{Phys. Lett. B \textbf{582} (2004) 237}}, \href{https://arxiv.org/abs/hep-th/0311240}{[arXiv:hep-th/0311240]}.

\bibitem{gbentropy1}
Mirjam Cvetic, Shin'ichi Nojiri, Sergei D. Odintsov,  ``Black Hole Thermodynamics and Negative Entropy in De Sitter and Anti-De Sitter Einstein-Gauss-Bonnet Gravity",
{\hypersetup{urlcolor=vividviolet}\href{https://www.sciencedirect.com/science/article/abs/pii/S0550321302000755?via\%3Dihub}{Nucl. Phys. B   \textbf{628} (2002)  295}}, \href{https://arxiv.org/abs/hep-th/0112045}{[arXiv:hep-th/0112045]}.


\bibitem{gbentropy2}
Tim Clunan, Simon F. Ross, Douglas J. Smith,  ``On Gauss-Bonnet Black Hole Entropy",
{\hypersetup{urlcolor=vividviolet}\href{https://iopscience.iop.org/article/10.1088/0264-9381/21/14/009}{ Class. Quantum  Grav.   \textbf{21} (2004)  3447}}, \href{https://arxiv.org/abs/gr-qc/0402044v4}{[arXiv:gr-qc/0402044]}.

\bibitem{gbentropy3}
Safia Taj, Hernando Quevedo, Alberto Sanchez,  ``Geometrothermodynamics of Five Dimensional Black Holes in Einstein–Gauss–Bonnet Theory",
{\hypersetup{urlcolor=vividviolet}\href{https://link.springer.com/article/10.1007\%2Fs10714-012-1351-6}{ Gen. Rel. Grav.   \textbf{44} (2012)  1489}}, \href{https://arxiv.org/abs/1104.3195}{[arXiv:math-ph/1104.3195]}.

\bibitem{0308056}
Roberto Emparan, Robert C. Myers, ``Instability of Ultra-Spinning Black Holes'', {\hypersetup{urlcolor=vividviolet}\href{https://doi.org/10.1088/1126-6708/2003/09/025}{JHEP \textbf{09} (2003) 025}}, \href{https://arxiv.org/abs/hep-th/0308056}{[arXiv:hep-th/0308056]}.

\bibitem{1412.5432}
Michael R.R. Good, Yen Chin Ong, ``Are Black Holes Springlike?'',  {\hypersetup{urlcolor=vividviolet}\href{https://journals.aps.org/prd/abstract/10.1103/PhysRevD.91.044031}{Phys. Rev. D \textbf{91} (2015) 044031}}, \href{https://arxiv.org/abs/1412.5432}{[arXiv:1412.5432 [gr-qc]]}.

\bibitem{1809.00442}
Yen Chin Ong, ``GUP-Corrected Black Hole Thermodynamics and the Maximum Force Conjecture'', {\hypersetup{urlcolor=vividviolet}\href{https://www.sciencedirect.com/science/article/pii/S0370269318306828?via\%3Dihub}{Phys. Lett. B \textbf{785} (2018) 217}}, \href{https://arxiv.org/abs/1809.00442}{[arXiv:1809.00442 [gr-qc]]}.


\bibitem{1803.03271}
Vitor Cardoso, Taishi Ikeda, Christopher J. Moore, Chul-Moon Yoo, ``Remarks on the Maximum Luminosity'', {\hypersetup{urlcolor=vividviolet}\href{https://journals.aps.org/prd/abstract/10.1103/PhysRevD.97.084013}{Phys. Rev. D \textbf{97} (2018) 084013}}, \href{https://arxiv.org/abs/1803.03271}{[arXiv:1803.03271 [gr-qc]]}.

\bibitem{2105.06650}
Aden Jowsey, Matt Visser, ``Reconsidering Maximum Luminosity'',  {\hypersetup{urlcolor=vividviolet}\href{https://www.worldscientific.com/doi/abs/10.1142/S0218271821420268}{Int. J. Mod. Phys. D \textbf{30} (2021) 14, 2142026}}, \href{https://arxiv.org/abs/2105.06650}{[arXiv:2105.06650 [gr-qc]]}.

\bibitem{2109.05973}
Li-Ming Cao, Long-Yue Li, Liang-Bi Wu, ``A Bound on the Rate of Bondi Mass Loss'', {\hypersetup{urlcolor=vividviolet}\href{https://journals.aps.org/prd/abstract/10.1103/PhysRevD.104.124017}{Phys. Rev. D \textbf{104} (2021) 12, 124017}}, \href{https://arxiv.org/abs/2109.05973}{[arXiv:2109.05973 [gr-qc]]}.

\bibitem{rup}
George Ruppeiner, ``Riemannian Geometry in Thermodynamic Fluctuation Theory'', {\hypersetup{urlcolor=vividviolet}\href{https://journals.aps.org/rmp/abstract/10.1103/RevModPhys.67.605}{Rev. Mod. Phys. \textbf{67} (1995) 605}}; Erratum: {\hypersetup{urlcolor=vividviolet}\href{https://journals.aps.org/rmp/abstract/10.1103/RevModPhys.68.313}{Rev. Mod. Phys. \textbf{68} (1996) 313}}.

\bibitem{1007.2160}
George Ruppeiner, ``Thermodynamic Curvature Measures Interactions'', {\hypersetup{urlcolor=vividviolet}\href{https://aapt.scitation.org/doi/10.1119/1.3459936}{Amer. Jour. Phys \textbf{78} (2010) 1170}}, \href{https://arxiv.org/abs/1007.2160}{[arXiv:1007.2160 [cond-mat.stat-mech]]}.


\bibitem{0802.1326}
George Ruppeiner, ``Thermodynamic Curvature and Phase Transitions in Kerr-Newman Black Holes'', {\hypersetup{urlcolor=vividviolet}\href{https://journals.aps.org/prd/abstract/10.1103/PhysRevD.78.024016}{Phys. Rev. D \textbf{78} (2008) 024016}}, \href{https://arxiv.org/abs/0802.1326}{[arXiv:0802.1326 [gr-qc]]}.


\bibitem{J1}
H. Janyszek, R. Mrugaa, ``Riemannian Geometry and Stability of Ideal Quantum Gases'', {\hypersetup{urlcolor=vividviolet}\href{https://iopscience.iop.org/article/10.1088/0305-4470/23/4/016}{J. Phys. A: Math. Gen. \textbf{23} (1990) 467}}.

\bibitem{J2}
H. Janyszek, ``Riemannian Geometry and Stability of Thermodynamical Equilibrium Systems'',  {\hypersetup{urlcolor=vividviolet}\href{https://iopscience.iop.org/article/10.1088/0305-4470/23/4/017}{J. Phys. A: Math. Gen. \textbf{23} (1990) 477}}.

\bibitem{1507.06097}
Jan E. {\AA}man, Ingemar Bengtsson, Narit Pidokrajt, ``Thermodynamic Metrics and Black Hole Physics'', {\hypersetup{urlcolor=vividviolet}\href{https://www.mdpi.com/1099-4300/17/9/6503}{Entropy \textbf{17} (2015) 6503}}, \href{https://arxiv.org/abs/1507.06097}{[arXiv:1507.06097 [gr-qc]]}.



\bibitem{331031a0}
John D. Barrow, Frank J. Tipler, ``Action Principles in Nature'', {\hypersetup{urlcolor=vividviolet}\href{https://www.nature.com/articles/331031a0}{Nature \textbf{331} (1988) 31}}.

\bibitem{1912.12926}
John D. Barrow, ``Finite Action Principle Revisited'', {\hypersetup{urlcolor=vividviolet}\href{https://journals.aps.org/prd/abstract/10.1103/PhysRevD.101.023527}{Phys. Rev. D \textbf{101} (2020) 023527}}, \href{https://arxiv.org/abs/1912.12926}{[arXiv:1912.12926 [gr-qc]]}.

\bibitem{1909.01169}
Jean-Luc Lehners, K.S. Stelle, ``A Safe Beginning for the Universe?'', {\hypersetup{urlcolor=vividviolet}\href{https://journals.aps.org/prd/abstract/10.1103/PhysRevD.100.083540}{Phys. Rev. D \textbf{100} (2019) 083540}}, \href{https://arxiv.org/abs/1909.01169}{[arXiv:1909.01169 [hep-th]]}.


\bibitem{2201.11132}
Raphael Bousso, Arvin Shahbazi-Moghaddam, ``Singularities From Entropy'', \href{https://arxiv.org/abs/2201.11132}{[arXiv:2201.11132 [hep-th]]}.

\bibitem{1103.3898}
Charis Anastopoulos, Ntina Savvidou, ``Entropy of Singularities in Self-Gravitating Radiation'', {\hypersetup{urlcolor=vividviolet}\href{https://iopscience.iop.org/article/10.1088/0264-9381/29/2/025004}{Class. Quant. Grav. \textbf{29} (2012) 025004}}, \href{https://arxiv.org/abs/1103.3898}{[arXiv:1103.3898 [gr-qc]]}.

\bibitem{2201.03096}
Raphael Bousso, Xi Dong, Netta Engelhardt, Thomas Faulkner, Thomas Hartman, Stephen H. Shenker, Douglas Stanford, ``Snowmass White Paper: Quantum Aspects of Black Holes and the Emergence of Spacetime'', \href{https://arxiv.org/abs/2201.03096}{[arXiv:2201.03096 [hep-th]]}.

\bibitem{2203.07117}
Thomas Faulkner, Thomas Hartman, Matthew Headrick, Mukund Rangamani, Brian Swingle, ``Snowmass White Paper: Quantum Information in Quantum Field Theory and Quantum Gravity'', \href{https://arxiv.org/abs/2203.07117}{[arXiv:2203.07117 [hep-th]]}.

\bibitem{caiqy}
Dongshan He, Qingyu Cai, ``Area Entropy and Quantized Mass of Black Holes from Information Theory", {\hypersetup{urlcolor=vividviolet}\href{https://www.mdpi.com/1099-4300/23/7/858}{Entropy \textbf{23} (2021) 7, 858}}.

\bibitem{1511.01162}
Ana Alonso-Serrano, Matt Visser, ``On Burning a Lump of Coal'', {\hypersetup{urlcolor=vividviolet}\href{https://www.sciencedirect.com/science/article/pii/S037026931630096X?via\%3Dihub}{Phys. Lett. B \textbf{757} (2016) 383}}, \href{https://arxiv.org/abs/1511.01162}{[arXiv:1511.01162 [gr-qc]]}.

\bibitem{1108.0417}
Anthony Aguirre, Sean M. Carroll, Matthew C. Johnson, ``Out of Equilibrium: Understanding Cosmological Evolution to Lower-Entropy States'', {\hypersetup{urlcolor=vividviolet}\href{https://iopscience.iop.org/article/10.1088/1475-7516/2012/02/024}{JCAP \textbf{02} (2012) 024}}, \href{https://arxiv.org/abs/1108.0417}{[arXiv:1108.0417 [hep-th]]}.

\bibitem{Dougherty}
John Dougherty, Craig Callender, ``Black Hole Thermodynamics: More Than an Analogy?'', PhilSci Archive, URL: \href{http://philsci-archive.pitt.edu/id/eprint/13195}{http://philsci-archive.pitt.edu/id/eprint/13195}. 

\bibitem{Helge}
Helge Kragh, ``Max Planck: The Reluctant Revolutionary'', Physics World, December 2020, URL: \href{https://physicsworld.com/a/max-planck-the-reluctant-revolutionary/}{https://physicsworld.com/a/max-planck-the-reluctant-revolutionary/}.

\bibitem{1511.06460}
Maulik Parikh, Andrew Svesko, ``Thermodynamic Origin of the Null Energy Condition'', {\hypersetup{urlcolor=vividviolet}\href{https://journals.aps.org/prd/abstract/10.1103/PhysRevD.95.104002}{Phys. Rev. D \textbf{95} (2017) 104002}}, \href{https://arxiv.org/abs/1511.06460}{[arXiv:1511.06460 [hep-th]]}.


\bibitem{2204.04352}
Michael te Vrugt, Paul Needham, Georg J. Schmitz, ``Is Thermodynamics Fundamental?'', \href{https://arxiv.org/abs/2204.04352}{[arXiv:2204.04352 [physics.hist-ph]]}.

\bibitem{9708200}
Elliott H. Lieb, Jakob Yngvason, ``The Physics and Mathematics of the Second Law of Thermodynamics'', {\hypersetup{urlcolor=vividviolet}\href{https://www.sciencedirect.com/science/article/abs/pii/S0370157398000829?via\%3Dihub}{Phys. Rept. \textbf{310} (1999) 1}}, \href{https://arxiv.org/abs/cond-mat/9708200}{[arXiv:cond-mat/9708200 [cond-mat.soft]]}.

\bibitem{9805005}
Elliott H. Lieb, Jakob Yngvason, ``A Guide to Entropy and the Second Law of Thermodynamics'', {\hypersetup{urlcolor=vividviolet}\href{https://www.ams.org/notices/199805/199805FullIssue.pdf}{Notices Amer. Math. Soc. \textbf{45} (1998) 571}}, \href{https://arxiv.org/abs/math-ph/9805005}{[arXiv:math-ph/9805005]}.


\bibitem{1205.2586}
Ovidiu Cristinel Stoica, ``Metric Dimensional Reduction at Singularities With Implications to Quantum Gravity'', {\hypersetup{urlcolor=vividviolet}\href{https://www.sciencedirect.com/science/article/abs/pii/S0003491614001080?via\%3Dihub}{Annals Phys. \textbf{347} (2014) 74}}, \href{https://arxiv.org/abs/1205.2586}{[arXiv:1205.2586 [gr-qc]]}.

\bibitem{1401.6283}
Ovidiu Cristinel Stoica, ``The Geometry of Black Hole Singularities'',  {\hypersetup{urlcolor=vividviolet}\href{https://www.hindawi.com/journals/ahep/2014/907518/}{Adv. High Energy Phys. \textbf{2014} (2014) 907518}}, \href{https://arxiv.org/abs/1401.6283}{[arXiv:1401.6283 [gr-qc]]}.

\bibitem{HawkingPenrose}
Stephen W. Hawking, Roger Penrose, \emph{The Nature of Space and Time}, Princeton Science Library, 2000.

\bibitem{0310281}
Gary T. Horowitz, Juan Maldacena, ``The Black Hole Final State'', {\hypersetup{urlcolor=vividviolet}\href{https://iopscience.iop.org/article/10.1088/1126-6708/2004/02/008}{JHEP \textbf{02} (2004) 008}}, \href{https://arxiv.org/abs/hep-th/0310281}{[arXiv:hep-th/0310281]}.

\bibitem{1911.02129}
Mariam Bouhmadi-L\'opez, Suddhasattwa Brahma, Che-Yu Chen, Pisin Chen, Dong-han Yeom, ``Annihilation-To-Nothing: A Quantum Gravitational Boundary Condition for the Schwarzschild Black Hole'', {\hypersetup{urlcolor=vividviolet}\href{https://iopscience.iop.org/article/10.1088/1475-7516/2020/11/002}{JCAP \textbf{11} (2020) 002}}, \href{https://arxiv.org/abs/1911.02129v2}{[arXiv:1911.02129 [gr-qc]]}.

\bibitem{2108.05744}
Malcolm J. Perry, ``Future Boundaries and the Black Hole Information Paradox'', \href{https://arxiv.org/abs/2108.05744}{[arXiv:2108.05744 [hep-th]]}.

\bibitem{0806.3818}
Brett McInnes, ``Black Hole Final State Conspiracies'', {\hypersetup{urlcolor=vividviolet}\href{https://www.sciencedirect.com/science/article/abs/pii/S0550321308004392?via\%3Dihub}{Nucl. Phys. B \textbf{807} (2009) 33}}, \href{https://arxiv.org/abs/0806.3818}{[arXiv:0806.3818 [hep-th]]}.

\bibitem{1710.07645}
Ludwig Eglseer, Stefan Hofmann, Marc Schneider, ``Quantum Populations Near Black-Hole Singularities'', {\hypersetup{urlcolor=vividviolet}\href{https://journals.aps.org/prd/abstract/10.1103/PhysRevD.104.105010}{Phys. Rev. D \textbf{104} (2021) 105010}}, \href{https://arxiv.org/abs/1710.07645}{[arXiv:1710.07645 [hep-th]]}.

\bibitem{1504.05580}
Stefan Hofmann, Marc Schneider, ``Classical Versus Quantum Completeness'', {\hypersetup{urlcolor=vividviolet}\href{https://journals.aps.org/prd/abstract/10.1103/PhysRevD.91.125028}{Phys. Rev. D \textbf{91} (2015) 125028}}, \href{https://arxiv.org/abs/1504.05580}{[arXiv:1504.05580 [hep-th]]}.

\bibitem{2106.03715}
Malcolm J. Perry, ``No Future in Black Holes'', \href{https://arxiv.org/abs/2106.03715}{[arXiv:2106.03715 [hep-th]]}.

\end{thebibliography}
\end{document}